\documentclass[twocolumn,prb,superscriptaddress,longbibliography]{revtex4-2}
\pdfoutput=1
\usepackage{graphicx}
\usepackage{color}
\usepackage{amsmath}
\usepackage{enumitem}
\usepackage{amssymb}
\usepackage[normalem]{ulem}
\usepackage{adjustbox}
\usepackage{bm}
\usepackage{braket}
\usepackage{physics}
\usepackage{dcolumn}
\usepackage{hyperref}
\usepackage{siunitx}
\usepackage{pdfpages}
\makeatletter
\patchcmd{\@outputpage@head}{\@ifx{\LS@rot\@undefined}{}{\LS@rot}}{}{}{}
\makeatother

\hypersetup{
	colorlinks = true,
	linkcolor = blue,
	citecolor = blue,
	urlcolor  = blue,
}

\DeclareMathAlphabet{\mathdutchcal}{U}{dutchcal}{m}{n}
\SetMathAlphabet{\mathdutchcal}{bold}{U}{dutchcal}{b}{n}
\DeclareMathAlphabet{\mathdutchbcal}{U}{dutchcal}{b}{n}

\begin{document}

\title{Twist-modulated magnetic interactions in bilayer van der Waals materials}

\author{Tomas T. Osterholt}
\affiliation{%
Institute for Theoretical Physics, Utrecht University, 3584CC Utrecht, The Netherlands\\
}%

\author{D. O. Oriekhov}
\affiliation{%
Kavli Institute of Nanoscience and QuTech, Delft University of Technology, 2628CJ Delft, The Netherlands\\
}%

\author{Lumen Eek}
\affiliation{%
Institute for Theoretical Physics, Utrecht University, 3584CC Utrecht, The Netherlands\\
}%

\author{Cristiane Morais Smith}
\affiliation{%
Institute for Theoretical Physics, Utrecht University, 3584CC Utrecht, The Netherlands\\
}

\author{Rembert A. Duine}
\affiliation{%
Institute for Theoretical Physics, Utrecht University, 3584CC Utrecht, The Netherlands\\
}
\affiliation{Department of Applied Physics, Eindhoven University of Technology,
P.O. Box 513, 5600 MB Eindhoven, The Netherlands
}%

\date{July 31, 2025}

\begin{abstract}
The ability to control magnetic interactions at the nanoscale is crucial for the development of next-generation spintronic devices and functional magnetic materials. In this work, we investigate theoretically, by means of many-body perturbation theory, how interlayer twisting modulates magnetic interactions in bilayer van der Waals systems composed of two ferromagnetic layers. We demonstrate that the relative strengths of the interlayer Heisenberg exchange interaction, the Dzyaloshinskii-Moriya interaction, and the anisotropic exchange interaction can be significantly altered by varying the twist angle between the layers, thus leading to tunable magnetic textures. We further show that these interactions are strongly dependent on the chemical potential, enabling additional control via electrostatic gating or doping. Importantly, our approach is applicable to arbitrary twist angles and does not rely on the construction of a Moiré supercell, making it particularly efficient even at small twist angles.
\end{abstract}

\maketitle

\textit{Introduction}. Two-dimensional (2D) materials have garnered significant attention due to their unique electronic \cite{Novoselov2005,Zhang2005,Goerbig2011,Forsythe2018,Dutta2024}, optical \cite{Falkovsky2008,Bernardi2017,Ketolainen2020}, and mechanical properties \cite{Papageorgiou2017,Akinwande2017,Kim2019}, which have promising applications in fields such as nanoelectronics \cite{Akinwande2014,Liu2018}, photonics \cite{Zhai2020}, and quantum computing \cite{Liu2019}. One particularly interesting subfield in this research domain is the field of twistronics \cite{Carr2017,Hennighausen2021,Jorio2022}, the study of how the electronic properties of layered 2D materials change with the relative twist angle between the layers. A striking example demonstrating the importance of this subfield is the discovery of unconventional superconductivity in twisted bilayer graphene at a `magic angle' of $1.1$ degrees \cite{Cao2018}, which was theoretically predicted by Bistritzer and MacDonald in 2011 \cite{Bistritzer2011}. 

Beyond electronic properties, the magnetic interactions in 2D materials have also become a topic of intense investigation \cite{Hallal2021,Cui2022,Elahi2024}. Because the formation of a particular magnetic texture is critically dependent on the relative strength of the different magnetic interactions, the ability to control these interactions becomes crucially important for both fundamental and practical applications. For example, magnetic skyrmions \cite{Bogdanov1989,Robler2006}, which hold great promise for spintronic logic and memory devices \cite{Fert2013,Zhou2014,Zhang2015}, require a Dzyaloshinskii-Moriya interaction (DMI) that is sufficiently large compared to the Heisenberg exchange interaction and anisotropy. 


This growing emphasis on tunability has motivated a number of studies aimed at understanding how the twist angle modulates interlayer magnetic interactions \cite{Sivadas2018,Akram2021,Xiao2021,Kim2023}.
These papers predominantly consist of first-principles (DFT) calculations on specific materials, most notably chromium(III)iodide ($\mathrm{CrI}_3$). While these studies have provided valuable insights into the stacking-dependent behavior of Heisenberg exchange interactions, DMI, and anisotropic exchange interactions, they are inherently material-specific and computationally intensive. In particular, the need to construct large moiré supercells limits their applicability to relatively large twist angles, making them ill-suited for systematically exploring the small-angle regime, where moiré effects are most pronounced.

So far, few attempts have been made to develop analytical or semi-analytical models that capture the essential physics of twist-angle dependent magnetic interactions. For example, while there are models \cite{Tong2018,Hejazi2020,Ray2021} that address the twist-angle dependence of bilayer magnetic textures, they typically treat the interlayer magnetic couplings as input parameters that must be obtained from experiments or first-principles calculations. 

Here, we develop a general theoretical framework in which these interlayer magnetic couplings are calculated as a function of both the twist angle and the chemical potential by means of analytical methods. Our approach is based on a perturbative calculation of the many-body ground state energy $E_0$ of the magnetic bilayer, which accurately describes the system in the limit of low temperatures. The essential physical content can then be captured in the form of a single integral expression, whose numerical complexity does not increase with decreasing twist angle $\theta$. Focusing specifically on the interlayer Heisenberg exchange, DMI, and anisotropic exchange interactions, we then demonstrate with this approach that suitable combinations of twisting and electrostatic gating or doping can sharply enhance one magnetic interaction relative to the others, thereby opening a pathway toward tunable magnetic materials.

\begin{figure}
    \centering
    \includegraphics[width=1\linewidth]{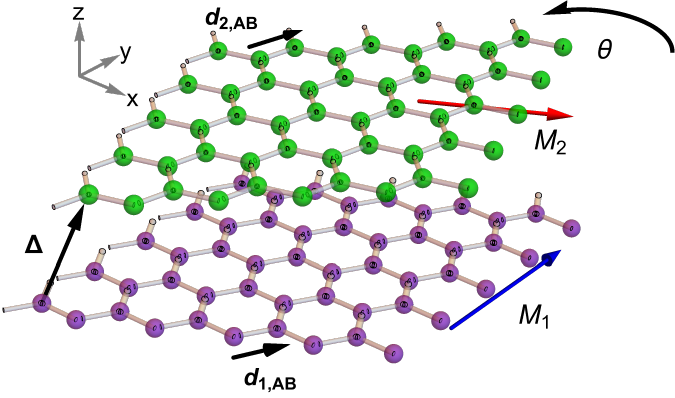}
    \caption{Two ferromagnetic layers with a honeycomb lattice structure stacked at a relative twist angle $\theta$ around the $z$-axis and a relative displacement $\boldsymbol{\Delta}$. Also shown are the exchange fields, $\mathbf{M}_1$ and $\mathbf{M}_2$, as well as the intralayer sublattice displacement vectors, $\mathbf{d}_{1,AB}$ and $\mathbf{d}_{2,AB}$, which connect the $A$- and $B$-sublattices within each layer.}
    \label{Figure Bilayer System Paper}
\end{figure}

\textit{Theory}. We consider a bilayer system of 2D crystalline ferromagnetic layers, which do not need to be of the same material. Each single layer consists of a finite number of equivalent Bravais sublattices that are translated with respect to one another. In a more mathematical language, denoting by $\{\mathbf{R}_{\alpha \gamma}\}$ and $\{\mathbf{R}_{\alpha \gamma'}\}$ the sets of lattice sites in the sublattices $\gamma$ and $\gamma'$ of the layer $\alpha$, we require that $\{\mathbf{R}_{\alpha \gamma'}\} = \{\mathbf{R}_{\alpha \gamma}+\mathbf{d}_{\alpha\gamma \gamma'}\} $. As illustrated in Fig. \ref{Figure Bilayer System Paper}, we allow for the layers to be translated by a vector $\mathbf{\Delta} = (\Delta_x,\Delta_y,\Delta_z)$ and rotated around the $z$-axis by an angle $\theta$ with respect to one another.

To describe the physics of such a bilayer system, we use a minimal model based on a mean-field approach. Within a single ferromagnetic layer $\alpha$, the electrons can hop to neighboring lattice sites $\mathbf{R}_{\alpha \gamma}$ and interact with the exchange field $\mathbf{M}_{\alpha}$ of said layer. To include interactions between the layers, we allow for interlayer electron hopping as well.

More formally, the Hamiltonian of the system takes the form $\hat{\mathcal{H}} = \hat{\mathcal{H}}_1 + \hat{\mathcal{H}}_2 + \hat{\mathcal{H}}_{\mathrm{int}}$,
where, introducing the shorthand notation $\eta = (\gamma,\sigma)$, we have
\begin{align}
    \hat{\mathcal{H}}_{\alpha} &= \sum_{\gamma,\gamma'}\sum_{\mathbf{R}_{\alpha \gamma},\mathbf{R}_{\alpha\gamma'}'} \sum_{\sigma,\sigma'} t^{(\alpha)}_{\eta \eta'} (\mathbf{R}_{\alpha\gamma}- \mathbf{R}_{\alpha \gamma'}')\hat{b}^{\dagger}_{ \sigma} (\mathbf{R}_{\alpha \gamma}) \hat{b}_{\sigma'}(\mathbf{R}_{\alpha\gamma'}') \nonumber \\& - g_{\alpha} \mathbf{M}_{\alpha} \cdot\sum_{\gamma}\sum_{\mathbf{R}_{\alpha\gamma}} \sum_{\sigma,\sigma'} \hat{b}^{\dagger}_{\sigma} (\mathbf{R}_{\alpha\gamma}) \mathbf{S}_{\sigma \sigma'} \hat{b}_{ \sigma'}(\mathbf{R}_{\alpha\gamma}), \nonumber \\
    \hat{\mathcal{H}}_{\mathrm{int}} &= \sum_{\gamma,\gamma'}\sum_{\mathbf{R}_{1\gamma},\mathbf{R}_{2\gamma'}}  \sum_{\sigma,\sigma'}  t^{\mathrm{int}}_{\eta \eta'}(\mathbf{R}_{1\gamma}-\mathbf{R}_{2\gamma'}) \hat{b}^{\dagger}_{\sigma}(\mathbf{R}_{1\gamma})  \hat{b}_{\sigma'}(\mathbf{R}_{2\gamma'}) \nonumber \\&+ \mathrm{h.c.}
\end{align}
In these expressions, $\hat{b}^{\dagger}_{\sigma}(\mathbf{R}_{\alpha\gamma})$ and $\hat{b}_{\sigma}(\mathbf{R}_{\alpha\gamma})$ are operators that respectively create and annihilate an electron with spin $\sigma$ at site $\mathbf{R}_{\alpha \gamma}$ in the Bravais sublattice $\gamma$ of the layer $\alpha$. We have also introduced the spin- and sublattice-dependent intralayer and interlayer tight-binding coefficients $t^{(\alpha)}_{\eta \eta'}$ and $t^{\mathrm{int}}_{\eta \eta'}$, the real constant $g_{\alpha} > 0$ that describes the strength of the coupling of the electrons to the exchange field $\mathbf{M}_{\alpha}$, and the vector of Pauli matrices $\mathbf{S} = (\sigma_x,\sigma_y,\sigma_z)$.

It is convenient to introduce the momentum creation and annihilation operators $\hat{b}^{\dagger}_{\gamma\sigma}(\mathbf{k}_{\alpha})$ and $\hat{b}_{\gamma\sigma}(\mathbf{k}_{\alpha})$ via a Fourier transformation. 
Assuming that the number of sublattices in the layer $\alpha$ is given by $m_{\alpha}$, one finds that the single-layer Hamiltonian $\hat{\mathcal{H}}_{\alpha}$ has $2m_{\alpha}$ energy bands $\epsilon_{\alpha,n s}(\mathbf{k}_{\alpha})$, with $n \in \{1, \cdots,m_{\alpha} \}$ and $s = \pm$, and that it can be expressed in the following diagonal form,
\begin{align}
    \hat{\mathcal{H}}_{\alpha} = \sum_{\{ns\}} \sum_{\mathbf{k}_{\alpha}} \epsilon_{\alpha,ns} (\mathbf{k}_{\alpha})\,  \hat{c}^{\dagger}_{n s}(\mathbf{k}_{\alpha}) \, \hat{c}_{n s}(\mathbf{k}_{\alpha}),
\end{align}
where the summations over $\mathbf{k}_1$ and $\mathbf{k}_2$ are restricted to the first Brillouin zone of respectively the first and second layer. The operators $\hat{c}^{\dagger}_{n s}(\mathbf{k}_{\alpha})$ and $\hat{c}_{n s}(\mathbf{k}_{\alpha})$ are related to $\hat{b}^{\dagger}_{\gamma\sigma}(\mathbf{k}_{\alpha})$ and $\hat{b}_{\gamma\sigma}(\mathbf{k}_{\alpha})$ via the following transformation,
\begin{align}
    \begin{cases}
    \hat{b}^{\dagger}_{\gamma \sigma} (\mathbf{k}_{\alpha}) &= \sum_{\{ns\}} \biggr[ U_{\alpha}^{\dagger}(\mathbf{k}_{\alpha}) \biggr]_{ns,\gamma \sigma} \hat{c}^{\dagger}_{n s}(\mathbf{k}_{\alpha}) ,
    \\ \hat{b}_{\gamma \sigma} (\mathbf{k}_{\alpha}) &= \sum_{\{ns\}} \biggr[ U_{\alpha}(\mathbf{k}_{\alpha})\biggr]_{\gamma\sigma,ns} \hat{c}_{n s}(\mathbf{k}_{\alpha}),
    \end{cases}
\end{align}
where the unitary $2 m_{\alpha} \times 2 m_{\alpha}$ matrix $U_{\alpha}(\mathbf{k}_{\alpha})$ satisfies
\begin{align}
    \biggr[ U_{\alpha}^{\dagger}(\mathbf{k}_{\alpha}) \mathcal{H}_{\alpha}(\mathbf{k}_{\alpha}) U_{\alpha}(\mathbf{k}_{\alpha})\biggr]_{n s,n's'} = \epsilon_{\alpha,ns}(\mathbf{k}_{\alpha}) \delta_{ns,n's'}.
\end{align}
Here, $ \mathcal{H}_{\alpha}(\mathbf{k}_{\alpha})$ is the single-particle Bloch Hamiltonian of the layer $\alpha$, which is described in detail in the Supplementary Materials (SM). 

Before moving on to discuss the interlayer Hamiltonian, we emphasize that the single-layer Hamiltonian $\hat{\mathcal{H}}_{\alpha}$ used here is the simplest possible Hamiltonian that leads to interlayer Heisenberg exchange, DMI and anisotropic exchange interactions. That being said, the methodology developed below applies equally well to more complicated forms of $\hat{\mathcal{H}}_{\alpha}$.

Now, just like the single-layer Hamiltonian $\hat{\mathcal{H}}_{\alpha}$, the interlayer Hamiltonian is also best understood in the momentum space representation \cite{Koshino2015}. Introducing the Fourier transform $\mathcal{T}^{\mathrm{int}}_{\eta \eta'}(\mathbf{k})$ of the interlayer tight-binding coefficient,
we find that the interlayer hopping Hamiltonian $\hat{\mathcal{H}}_{\mathrm{int}}$ can be written as
\begin{align}
    \hat{\mathcal{H}}_{\mathrm{int}} &= \sum_{\gamma, \gamma'} \sum_{\sigma, \sigma'} \sum_{\mathbf{k}_1,\mathbf{k}_2} \Gamma_{\eta\eta'}^{\mathrm{int}}(\mathbf{k}_1,\mathbf{k}_2) \hat{b}_{\gamma \sigma}^{\dagger}(\mathbf{k}_1) \hat{b}_{\gamma'\sigma'}(\mathbf{k}_2) + \, \mathrm{h.c.},
\end{align}
with
\begin{align}\label{Equation Gamma Paper}
    \Gamma_{\eta\eta'}^{\mathrm{int}}(\mathbf{k}_1,\mathbf{k}_2) &= \frac{1}{\sqrt{\mathcal{A}_{1}\mathcal{A}_{2}}} \sum_{\mathbf{G}_1,\mathbf{G}_2} \mathcal{T}^{\mathrm{int}}_{\eta \eta'}(\mathbf{G}_1-\mathbf{k}_1) e^{-i\mathbf{G}_2 \cdot \mathbf{\Delta}} \nonumber \\&\times \delta_{\mathbf{k}_1-\mathbf{G}_1,\mathbf{k}_2-\mathbf{G}_2}.
\end{align}
Here, $\mathcal{A}_1$ and $\mathcal{A}_2$ are, respectively, the areas of a primitive unit cell of a Bravais sublattice in the first and in the second layer, while $\mathbf{G}_1$ and $\mathbf{G}_2$ are, respectively, the reciprocal lattice vectors corresponding to the Bravais sublattices of the first and the second layer. Physically, this equation tells us that the interlayer Hamiltonian couples an electronic state with momentum $\mathbf{k}_1$ in the first layer to an electronic state with momentum $\mathbf{k}_2$ in the second layer only if the Umklapp scattering condition $\mathbf{k}_1-\mathbf{G}_1 = \mathbf{k}_2 -\mathbf{G}_2$ is satisfied \cite{Koshino2015}.


\begin{figure*}[t]
    \centering
    \includegraphics[scale=1]{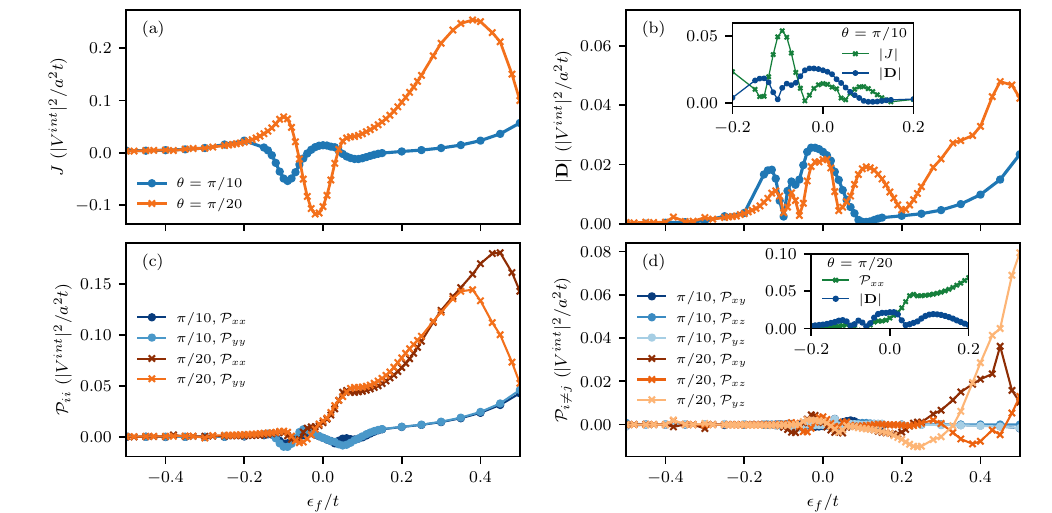}
    \caption{Plots of the interlayer Heisenberg exchange constant $J$, the magnitude of the  Dzyaloshinskii-Moriya vector $|\mathbf{D}|$, and the five independent components of the traceless, symmetric anistropy matrix $\mathcal{P}$ as functions of the Fermi energy $\epsilon_f$, for twist angles $\theta = \pi/10$ ($18^{\circ}$) and $\theta = \pi/20$ ($9^{\circ}$). The interlayer hopping is of the Slater-Koster type, $t^{\mathrm{int}}_{\sigma \sigma'}(\mathbf{r}) = V^{\mathrm{int}}  \exp(- r/r_0) \delta_{\sigma \sigma'}$, with $r_0$ the decay length and $V^{\mathrm{int}}$ the interaction energy constant.  The following parameters were used: $t = 1 $ (arb. units), $t_{\uparrow \downarrow} = 0.1 t$, $a=\Delta_z = r_0 = 1$ (arb. units), $\Delta_x = \Delta_y = 0$. (a) The interlayer Heisenberg exchange constant $J$. (b) The magnitude of the interlayer Dzyaloshinskii-Moriya vector $|\mathbf{D}|$. (c) The independent diagonal and (d) off-diagonal components of the interlayer anisotropic exchange matrix $\mathcal{P}$. }
    \label{Figure Results Paper 1}
\end{figure*}

\begin{figure}
    \centering
    \includegraphics[scale=1]{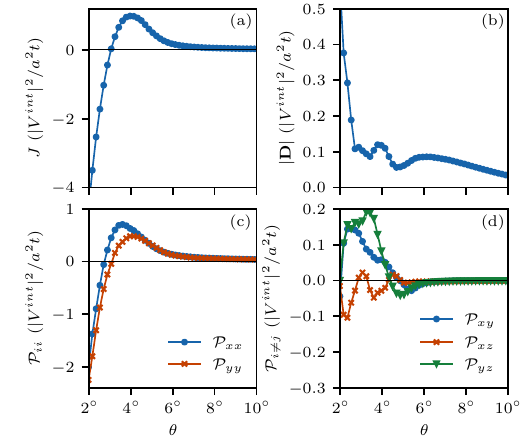}
    \caption{Plots of the interlayer Heisenberg exchange constant $J$, the magnitude of the  Dzyaloshinskii-Moriya vector $|\mathbf{D}|$, and the five independent components of the traceless, symmetric anistropy matrix $\mathcal{P}$ as functions of the twist angle $\theta$, for the Fermi energy $\epsilon_f = 0.2t$. The interlayer hopping is of the Slater-Koster type, $t^{\mathrm{int}}_{\sigma \sigma'}(\mathbf{r}) = V^{\mathrm{int}}  \exp(- r/r_0) \delta_{\sigma \sigma'}$.  The following parameters were used: $t = 1$ (arb. units), $t_{\uparrow \downarrow} = 0.1 t$, $a=\Delta_z = r_0 = 1$ (arb. units), $\Delta_x = \Delta_y = 0$. (a) The interlayer Heisenberg exchange constant $J$. (b) The magnitude of the interlayer Dzyaloshinskii-Moriya vector $|\mathbf{D}|$. (c) The independent diagonal and (d) off-diagonal components of the interlayer anisotropic exchange matrix $\mathcal{P}$. }
    \label{Figure Results Paper 2}
\end{figure}

With these momentum space representations at hand, we can now use perturbation theory around the non-interacting Hamiltonian $\hat{\mathcal{H}}_1+\hat{\mathcal{H}}_2$, treating $\hat{\mathcal{H}}_{\mathrm{int}}$ as a perturbation. The many-body ground state energy $E_0$ of $\hat{\mathcal{H}}$ can then be expressed as a perturbative series in $\hat{\mathcal{H}}_{\mathrm{int}}$, where we write $E_0 = E_0^{(0)} + E_0^{(1)} + E_0^{(2)} + \,\cdots$.
Since the first-order correction $E_0^{(1)}$ vanishes, we will only consider the second-order correction $E^{(2)}_0$ here. By introducing the matrix $\Upsilon^{\mathrm{int}}(\mathbf{k}_1,\mathbf{k}_2) =U_1^{\dagger}(\mathbf{k}_1) \,\Gamma^{\mathrm{int}}(\mathbf{k}_1,\mathbf{k}_2) U_2(\mathbf{k}_2)$,
we find that the second-order correction to the ground state energy is given by
\begin{widetext} 
\begin{align}\label{Equation Second-Order Correction Paper}
     E_0^{(2)} = \sum_{\{ns\}} \sum_{\{n's'\}}\sum_{\mathbf{k}_1,\mathbf{k}_2} \bigg| \Upsilon^{\mathrm{int}}_{ns,n's'} (\mathbf{k}_1,\mathbf{k}_2) \bigg|^2 \frac{F(\epsilon_{2,n's'}(\mathbf{k}_2))-F(\epsilon_{1,ns}(\mathbf{k}_1))}{\epsilon_{2,n's'}(\mathbf{k}_2)-\epsilon_{1,ns}(\mathbf{k}_1)}.
\end{align}
\end{widetext}
Here, we have introduced the zero-temperature Fermi-Dirac distribution functions $F(\epsilon) = 1-H(\epsilon-\epsilon_f)$, with the chemical potential set at the Fermi energy $\epsilon_f$ of the unperturbed system and where $H(\epsilon)$ denotes the Heaviside step function. We note that the twist-angle dependence in Eq.~\eqref{Equation Second-Order Correction Paper} originates from the relative rotation of the two Brillouin zones, which affects both Umklapp scattering ---via rotated reciprocal lattice vectors --- and the single-layer band energies through the rotation of the underlying dispersion.

To evaluate $E^{(2)}_0$ numerically, one can use the fact that $\mathcal{T}^{\mathrm{int}}(\mathbf{k})$ typically decreases rapidly for large $|\mathbf{k}|$, which allows one to restrict the summation over $\mathbf{G}_1$ and $\mathbf{G}_2$ to only the first few shortest reciprocal lattice vectors \cite{Catarina2019}. Furthermore, in the thermodynamic limit, we can make the replacement $\sum_{\mathbf{k}_{\alpha}} \rightarrow \mathcal{A}\int_{\Omega_{\alpha}} d^2 \mathbf{k}_{\alpha}/(2\pi)^2 $, where $\mathcal{A}$ is the total area of a single ferromagnetic layer and $\Omega_{\alpha}$ is the first Brillouin zone of the layer $\alpha$.

Finally, we note that the interlayer Heisenberg exchange energy density, $J(\theta)\,\mathbf{M}_1 \cdot \mathbf{M}_2$, the Dzyaloshinskii-Moriya energy density, $\mathbf{D}(\theta) \cdot (\mathbf{M}_1 \times \mathbf{M}_2)$, and the anisotropic exchange energy density, $\mathbf{M}_1^T \, \mathcal{P}(\theta) \, \mathbf{M}_2$, can be determined from the second-order ground-state energy correction. This can be done by using the fact that $E^{(2)}_0$ can be Taylor expanded in the vector components $M_{1,p}$ and $M_{2,q}$ of the layer exchange fields, thereby allowing us to extract the values of the Heisenberg exchange constant $J(\theta)$, the Dzyaloshinskii-Moriya vector $\mathbf{D}(\theta)$ and the anisotropic exchange matrix $\mathcal{P}(\theta)$ in a straightforward manner. The details of this extraction procedure can be found in the SM.

\textit{Results}. To demonstrate the strong twist-angle and chemical potential dependence of the interlayer magnetic interactions, we now apply our theoretical framework to the simplest possible bilayer system that we can consider. We take both layers to be completely equivalent square lattices, with the only possible differences between them being that $\mathbf{M}_1$ and $\mathbf{M}_2$ are allowed to point in different directions and that the layers are allowed to be rotated and translated with respect to one another. Furthermore, we consider intralayer hopping to be only between nearest neighbors, and we assume that $t_{\uparrow \uparrow} = t_{\downarrow \downarrow} = t/2$. A Rashba-type spin-orbit coupling (SOC) is included via the spin-flip hopping coefficients $t_{\uparrow \downarrow}$ and $t_{\downarrow \uparrow}$. Finally, the interlayer hopping is assumed to be of the Slater-Koster form \cite{Slater1954}, and we only include  reciprocal lattice vectors of length $0$ or $2\pi/a$, with $a$ the lattice constant, in Eq.~\eqref{Equation Gamma Paper}. 

In Fig. \ref{Figure Results Paper 1}, we present the calculated values of $J$, $|\mathbf{D}|$ and the five independent components of the traceless, symmetric anisotropic exchange matrix $\mathcal{P}$ as functions of the Fermi energy $\epsilon_f$ (i. e. the zero-temperature chemical potential) for twist angles $\theta = \pi/10$ ($18 ^{\circ}$) and $\theta = \pi/20$ ($9 ^{\circ}$). As shown in the inset of Fig. \ref{Figure Results Paper 1}(b), for a twist angle $\theta = \pi/10$, there exist values of the chemical potential for which the interlayer DMI significantly exceeds the interlayer Heisenberg exchange interaction. The relative importance of the various magnetic couplings, however, depends sensitively on the twist angle. For example, while the anisotropic exchange is largely negligible at $\theta=\pi/10$, it becomes substantially more relevant --- particularly compared to the DMI --- at $\theta = \pi/20$, as illustrated in the inset of Fig. \ref{Figure Results Paper 1}(d). Interestingly, we also observe in Fig. \ref{Figure Results Paper 1}(a) that the interlayer exchange constant $J$ oscillates around zero as a function of the chemical potential, leading to alternating ferro- and antiferromagnetic regimes. Moreover, we notice that the sign of $J$ depends strongly on the twist angle, with the signs of $J$ for $\theta = \pi/10$ and $\theta = \pi/20$ being mostly opposite to one another in a narrow range around $\epsilon_f = 0.2t$. 

In Fig. \ref{Figure Results Paper 2}, we present the calculated values of $J$, $|\mathbf{D}|$ and the five independent components of the traceless, symmetric anisotropic exchange matrix $\mathcal{P}$ as functions of the twist angle $\theta$, ranging from $1^{\circ}$ to $10^{\circ}$, for $\epsilon_f = 0.2t$. In Fig.  \ref{Figure Results Paper 2}(a), we again observe that the interlayer Heisenberg exchange interaction changes sign depending on the particular angle $\theta$ that is considered. Furthermore, we observe that all interactions decrease in magnitude as $\theta$ is increased, as one would intuitively expect from the suppression of Umklapp scattering at larger twist angles. 

Additional figures provided in the SM show that the direction of the DMI vector $\mathbf{D}$ is also highly sensitive to both the twist angle and the chemical potential.

\textit{Conclusion}. Our results reveal that interlayer Heisenberg exchange, DMI, and anisotropic exchange are highly sensitive to variations in twist angle and chemical potential. Such sensitivity offers a powerful mechanism for controlling magnetic interactions in bilayer van der Waals systems, thereby opening new avenues for the development of tunable magnetic materials and devices. 

Perhaps even more importantly, we have developed a general theoretical framework for calculating these interlayer magnetic interactions. This method is computationally far less demanding than first-principles approaches and maintains its efficiency regardless of the twist angle magnitude. Additionally, further developments of such analytical methods can offer clear physical insights that are often difficult to extract from DFT-based approaches, thereby providing a valuable complementary tool in the rapidly evolving field of twistronics-based magnetism.

We expect our approach, which was based on many-body perturbation theory, to work well when the energy bands of the single-layer Hamiltonians are not flat close to the Fermi level and when one can safely assume that the interlayer interaction is weak compared to the energy scale of the unperturbed system. Although we have assumed in this work that only one orbital state is available per lattice site, our model can be generalized straightforwardly to include any number of orbitals.

Finally, while we have primarily focused on interlayer Heisenberg exchange, DMI, and anisotropic exchange, the framework that we have developed can readily be applied to explore a broader class of magnetic interactions, which can be accessed, for example, by modifying the single-layer Hamiltonian. Among such interactions, we note that biquadratic exchange \cite{Kartsev2020,Ni2021,Oriekhov2025} is of particular interest, as this exotic interaction was found to be the dominant interaction in chemical compounds like $\mathrm{NiCl}_2$ and $\mathrm{NiBr}_2$ \cite{Ni2021}. Other directions of research that can be explored within this framework include the twist-angle dependence of \textit{intralayer} exchange and DMI, and the additional effects of relative sliding between the layers on the magnetic couplings.

\textit{Acknowledgements}. The work of T.T.O. and R.A.D. was supported by the Dutch Research Council (NWO) by the research programme OCENW.XL21.XL21.058. D.O.O. acknowledges the support by the Kavli Foundation. L.E. and C.M.S. acknowledge the research program \textit{Materials for the Quantum Age} (QuMat) for financial support. This program (registration number 024.005.006) is part of the Gravitation program financed by the Dutch Ministry of Education, Culture and Science (OCW). The authors acknowledge the use of computational resources
 of the DelftBlue supercomputer, provided by Delft High Performance Computing Centre \cite{DHPC2024}.

{\it Data availability statement}
The code to reproduce research in this paper can be found at \cite{Supplement-code}.
 


\bibliography{biblio}

\onecolumngrid
\newpage
\includepdf[pages={1,{},2-22}]{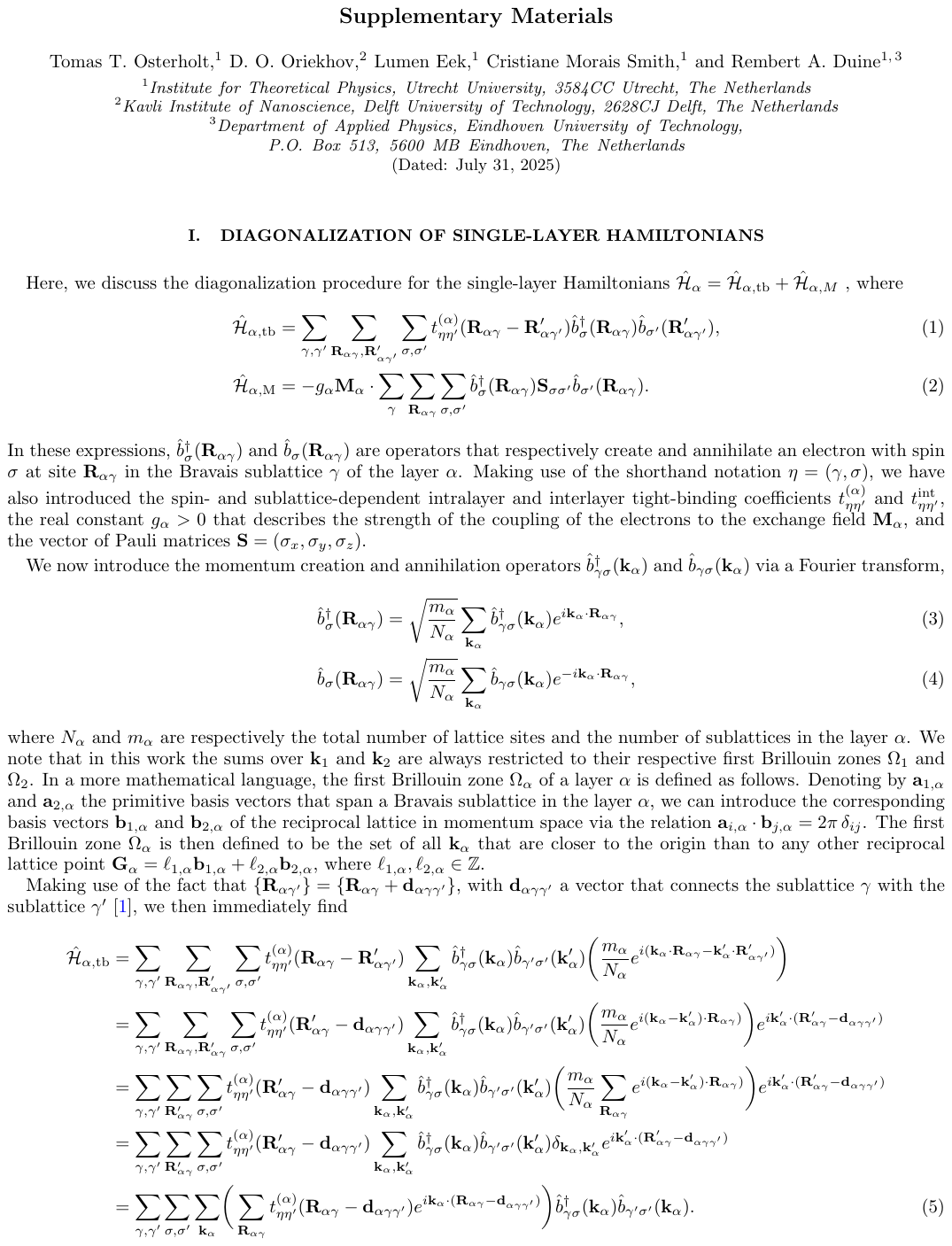}

\end{document}


\title{Supplementary Materials}

\author{Tomas T. Osterholt}
\affiliation{%
Institute for Theoretical Physics, Utrecht University, 3584CC Utrecht, The Netherlands\\
}%

\author{D. O. Oriekhov}
\affiliation{%
Kavli Institute of Nanoscience, Delft University of Technology, 2628CJ Delft, The Netherlands\\
}%

\author{Lumen Eek}
\affiliation{%
Institute for Theoretical Physics, Utrecht University, 3584CC Utrecht, The Netherlands\\
}%

\author{Cristiane Morais Smith}
\affiliation{%
Institute for Theoretical Physics, Utrecht University, 3584CC Utrecht, The Netherlands\\
}

\author{Rembert A. Duine}
\affiliation{%
Institute for Theoretical Physics, Utrecht University, 3584CC Utrecht, The Netherlands\\
}
\affiliation{Department of Applied Physics, Eindhoven University of Technology,
P.O. Box 513, 5600 MB Eindhoven, The Netherlands
}%

\date{July 31, 2025}

\maketitle   

\section{Diagonalization of single-layer Hamiltonians}
Here, we discuss the diagonalization procedure for the single-layer Hamiltonians $\hat{\mathcal{H}}_{\alpha} = \hat{\mathcal{H}}_{\alpha,\mathrm{tb}}+\hat{\mathcal{H}}_{\alpha,M}$ , where
\begin{align}
    \hat{\mathcal{H}}_{\alpha,\mathrm{tb}} &= \sum_{\gamma,\gamma'}\sum_{\mathbf{R}_{\alpha \gamma},\mathbf{R}_{\alpha\gamma'}'} \sum_{\sigma,\sigma'} t^{(\alpha)}_{\eta \eta'} (\mathbf{R}_{\alpha\gamma}- \mathbf{R}_{\alpha \gamma'}')\hat{b}^{\dagger}_{ \sigma} (\mathbf{R}_{\alpha \gamma}) \hat{b}_{\sigma'}(\mathbf{R}_{\alpha\gamma'}'), \\
    \hat{\mathcal{H}}_{\alpha,\mathrm{M}} &= - g_{\alpha} \mathbf{M}_{\alpha} \cdot\sum_{\gamma}\sum_{\mathbf{R}_{\alpha\gamma}} \sum_{\sigma,\sigma'} \hat{b}^{\dagger}_{\sigma} (\mathbf{R}_{\alpha\gamma}) \mathbf{S}_{\sigma \sigma'} \hat{b}_{ \sigma'}(\mathbf{R}_{\alpha\gamma}). 
\end{align}
In these expressions, $\hat{b}^{\dagger}_{\sigma}(\mathbf{R}_{\alpha\gamma})$ and $\hat{b}_{\sigma}(\mathbf{R}_{\alpha\gamma})$ are operators that respectively create and annihilate an electron with spin $\sigma$ at site $\mathbf{R}_{\alpha \gamma}$ in the Bravais sublattice $\gamma$ of the layer $\alpha$. Making use of the shorthand notation $\eta = (\gamma,\sigma)$, we have also introduced the spin- and sublattice-dependent intralayer and interlayer tight-binding coefficients $t^{(\alpha)}_{\eta \eta'}$ and $t^{\mathrm{int}}_{\eta \eta'}$, the real constant $g_{\alpha} > 0$ that describes the strength of the coupling of the electrons to the exchange field $\mathbf{M}_{\alpha}$, and the vector of Pauli matrices $\mathbf{S} = (\sigma_x,\sigma_y,\sigma_z)$.

We now introduce the momentum creation and annihilation operators $\hat{b}^{\dagger}_{\gamma \sigma}(\mathbf{k}_{\alpha})$ and $\hat{b}_{\gamma \sigma}(\mathbf{k}_{\alpha})$ via a Fourier transform,
\begin{align}
    \hat{b}^{\dagger}_{\sigma}(\mathbf{R}_{\alpha\gamma}) &= \sqrt{\frac{m_{\alpha}}{N_{\alpha}}} \sum_{\mathbf{k}_{\alpha}} \hat{b}^{\dagger}_{\gamma\sigma}(\mathbf{k}_{\alpha}) e^{i\mathbf{k}_{\alpha}\cdot\mathbf{R}_{\alpha \gamma}}, \\
    \hat{b}_{\sigma}(\mathbf{R}_{\alpha\gamma}) &= \sqrt{\frac{m_{\alpha}}{N_{\alpha}}} \sum_{\mathbf{k}_{\alpha}} \hat{b}_{\gamma\sigma}(\mathbf{k}_{\alpha}) e^{-i\mathbf{k}_{\alpha}\cdot\mathbf{R}_{\alpha \gamma}},
\end{align}
where $N_{\alpha}$ and $m_{\alpha}$ are respectively the total number of lattice sites and the number of sublattices in the layer $\alpha$. We note that in this work the sums over $\mathbf{k}_1$ and $\mathbf{k}_2$ are always restricted to their respective first Brillouin zones $\Omega_1$ and $\Omega_2$. In a more mathematical language, the first Brillouin zone $\Omega_{\alpha}$ of a layer $\alpha$ is defined as follows. Denoting by $\mathbf{a}_{1,\alpha}$ and $\mathbf{a}_{2,\alpha}$ the primitive basis vectors that span a Bravais sublattice in the layer $\alpha$, we can introduce the corresponding basis vectors $\mathbf{b}_{1,\alpha}$ and $\mathbf{b}_{2,\alpha}$ of the reciprocal lattice in momentum space via the relation $\mathbf{a}_{i,\alpha} \cdot \mathbf{b}_{j,\alpha} = 2\pi\,\delta_{ij}$. The first Brillouin zone $\Omega_{\alpha}$ is then defined to be the set of all $\mathbf{k}_{\alpha}$ that are closer to the origin than to any other reciprocal lattice point $\mathbf{G}_{\alpha} = \ell_{1,\alpha} \mathbf{b}_{1,\alpha} + \ell_{2,\alpha} \mathbf{b}_{2,\alpha} $, where $\ell_{1,\alpha},\ell_{2,\alpha} \in \mathbb{Z}$.

Making use of the fact that $\{\mathbf{R}_{\alpha\gamma'}\} = \{\mathbf{R}_{\alpha\gamma} + \mathbf{d}_{\alpha\gamma \gamma'} \}$, with $\mathbf{d}_{\alpha\gamma \gamma'}$ a vector that connects the sublattice $\gamma$ with the sublattice $\gamma'$ \footnote{To clarify this definition, one can consider the example of a honeycomb lattice. Suppose that we define our $A$- and $B$-sublattices such that we have an $A$-site at the origin and a $B$-site at $( a/\sqrt{3},0)$, with $a$ the lattice constant. We can then define $\mathbf{d}_{\alpha,AB} = a/\sqrt{3}\,\hat{\mathbf{x}}$, $\mathbf{d}_{\alpha,BA} = -a/\sqrt{3}\,\hat{\mathbf{x}}$, and $\mathbf{d}_{\alpha,AA} = \mathbf{d}_{\alpha,BB} = \mathbf{0}$.}, we then immediately find
\begin{align}
    \hat{\mathcal{H}}_{\alpha,\mathrm{tb}} &= \sum_{\gamma,\gamma'}\sum_{\mathbf{R}_{\alpha \gamma},\mathbf{R}_{\alpha\gamma'}'} \sum_{\sigma,\sigma'} t^{(\alpha)}_{\eta \eta'} (\mathbf{R}_{\alpha\gamma}- \mathbf{R}_{\alpha \gamma'}') \sum_{\mathbf{k}_{\alpha},\mathbf{k}_{\alpha}'}\hat{b}^{\dagger}_{ \gamma\sigma} (\mathbf{k}_{\alpha}) \hat{b}_{\gamma'\sigma'}(\mathbf{k}_{\alpha}')
     \biggr( \frac{m_{\alpha}}{N_{\alpha}} e^{i (\mathbf{k}_{\alpha} \cdot\mathbf{R}_{\alpha\gamma}-\mathbf{k}_{\alpha}'\cdot\mathbf{R}_{\alpha \gamma'}')} \biggr) \nonumber \\
    &= \sum_{\gamma,\gamma'}\sum_{\mathbf{R}_{\alpha \gamma},\mathbf{R}_{\alpha\gamma}'} \sum_{\sigma,\sigma'} t^{(\alpha)}_{\eta \eta'} (\mathbf{R}_{\alpha\gamma}'- \mathbf{d}_{\alpha \gamma \gamma'}) \sum_{\mathbf{k}_{\alpha},\mathbf{k}_{\alpha}'}\hat{b}^{\dagger}_{ \gamma\sigma} (\mathbf{k}_{\alpha}) \hat{b}_{\gamma'\sigma'}(\mathbf{k}_{\alpha}') \biggr( \frac{m_{\alpha}}{N_{\alpha}} e^{i (\mathbf{k}_{\alpha}-\mathbf{k}_{\alpha}')\cdot\mathbf{R}_{\alpha \gamma})} \biggr)e^{i \mathbf{k}_{\alpha}'\cdot(\mathbf{R}_{\alpha \gamma}'-\mathbf{d}_{\alpha \gamma \gamma'})} \nonumber\\
    &= \sum_{\gamma,\gamma'}\sum_{\mathbf{R}_{\alpha\gamma}'} \sum_{\sigma,\sigma'} t^{(\alpha)}_{\eta \eta'} (\mathbf{R}_{\alpha\gamma}'- \mathbf{d}_{\alpha \gamma \gamma'}) \sum_{\mathbf{k}_{\alpha},\mathbf{k}_{\alpha}'}\hat{b}^{\dagger}_{ \gamma\sigma} (\mathbf{k}_{\alpha}) \hat{b}_{\gamma'\sigma'}(\mathbf{k}_{\alpha}') \biggr( \frac{m_{\alpha}}{N_{\alpha}} \sum_{\mathbf{R}_{\alpha \gamma}} e^{i (\mathbf{k}_{\alpha}-\mathbf{k}_{\alpha}')\cdot\mathbf{R}_{\alpha \gamma})} \biggr)e^{i \mathbf{k}_{\alpha}'\cdot(\mathbf{R}_{\alpha \gamma}'-\mathbf{d}_{\alpha \gamma \gamma'})} \nonumber \\
    &= \sum_{\gamma,\gamma'}\sum_{\mathbf{R}_{\alpha\gamma}'} \sum_{\sigma,\sigma'} t^{(\alpha)}_{\eta \eta'} (\mathbf{R}_{\alpha\gamma}'- \mathbf{d}_{\alpha \gamma \gamma'}) \sum_{\mathbf{k}_{\alpha},\mathbf{k}_{\alpha}'}\hat{b}^{\dagger}_{ \gamma\sigma} (\mathbf{k}_{\alpha}) \hat{b}_{\gamma'\sigma'}(\mathbf{k}_{\alpha}') \delta_{\mathbf{k}_{\alpha},\mathbf{k}_{\alpha}'}e^{i \mathbf{k}_{\alpha}'\cdot(\mathbf{R}_{\alpha \gamma}'-\mathbf{d}_{\alpha \gamma \gamma'})} \nonumber \\
    &= \sum_{\gamma,\gamma'} \sum_{\sigma,\sigma'} \sum_{\mathbf{k}_{\alpha}} \biggr( \sum_{\mathbf{R}_{\alpha \gamma}} t^{(\alpha)}_{\eta \eta'} (\mathbf{R}_{\alpha\gamma}- \mathbf{d}_{\alpha \gamma \gamma'}) e^{i \mathbf{k}_{\alpha}\cdot(\mathbf{R}_{\alpha \gamma}-\mathbf{d}_{\alpha \gamma \gamma'})} \biggr)\hat{b}^{\dagger}_{ \gamma\sigma} (\mathbf{k}_{\alpha}) \hat{b}_{\gamma'\sigma'}(\mathbf{k}_{\alpha}).
\end{align}
In the second line we wrote $\mathbf{R}_{\alpha \gamma'}' = \mathbf{R}_{\alpha \gamma}-\mathbf{R}_{\alpha \gamma}'+\mathbf{d}_{\alpha \gamma \gamma'}$ and changed the summation from $\mathbf{R}_{\alpha \gamma'}'$ to $\mathbf{R}_{\alpha \gamma}'$. In the final line, we changed the dummy index from $\mathbf{R}_{\alpha\gamma}'$ to $\mathbf{R}_{\alpha\gamma}$. In a similar manner, we have
\begin{align}
    \hat{\mathcal{H}}_{\alpha,\mathrm{M}} &= - g_{\alpha} \mathbf{M}_{\alpha} \cdot \sum_{\gamma} \sum_{\sigma, \sigma'} \sum_{\mathbf{k}_{\alpha},\mathbf{k}_{\alpha}'} \hat{b}^{\dagger}_{\gamma \sigma} (\mathbf{k}_{\alpha}) \mathbf{S}_{\sigma \sigma'}\hat{b}_{\gamma \sigma'} (\mathbf{k}_{\alpha}') \biggr(\frac{m_{\alpha}}{N_{\alpha}} \sum_{\mathbf{R}_{\alpha\gamma}}e^{i (\mathbf{k}_{\alpha}-\mathbf{k}_{\alpha}')\cdot\mathbf{R}_{\alpha\gamma}}\biggr)\nonumber\\
    &= - g_{\alpha} \mathbf{M}_{\alpha} \cdot \sum_{\gamma} \sum_{\sigma, \sigma'} \sum_{\mathbf{k}_{\alpha}}  \hat{b}^{\dagger}_{\gamma \sigma} (\mathbf{k}_{\alpha})\mathbf{S}_{\sigma \sigma'}\hat{b}_{\gamma \sigma'} (\mathbf{k}_{\alpha}) \nonumber \\
    &=- g_{\alpha}\mathbf{M}_{\alpha} \cdot \sum_{\gamma,\gamma'} \sum_{\sigma, \sigma'} \sum_{\mathbf{k}_{\alpha}}  \delta_{\gamma\gamma'} \,\hat{b}^{\dagger}_{\gamma \sigma} (\mathbf{k}_{\alpha})\mathbf{S}_{\sigma \sigma'}\hat{b}_{\gamma' \sigma'} (\mathbf{k}_{\alpha}).
\end{align}
We can now straightforwardly write $\hat{\mathcal{H}}_{\alpha}$ in its diagonal basis,
\begin{align}
    \hat{\mathcal{H}}_{\alpha} = \sum_{\{ns\}} \sum_{\mathbf{k}} \epsilon_{\alpha,ns} (\mathbf{k}_{\alpha})\,  \hat{c}^{\dagger}_{n s}(\mathbf{k}_{\alpha}) \, \hat{c}_{n s}(\mathbf{k}_{\alpha}).
\end{align}
The operators $\hat{c}^{\dagger}_{n s}(\mathbf{k}_{\alpha})$ and $\hat{c}_{n s}(\mathbf{k}_{\alpha})$ are related to $\hat{b}^{\dagger}_{\gamma\sigma}(\mathbf{k}_{\alpha})$ and $\hat{b}_{\gamma\sigma}(\mathbf{k}_{\alpha})$ via the following transformation,
\begin{align}
    \hat{b}^{\dagger}_{\gamma \sigma} (\mathbf{k}_{\alpha}) &= \sum_{\{ns\}} \biggr( U_{\alpha}^{\dagger}(\mathbf{k}_{\alpha}) \biggr)_{ns,\gamma \sigma} \hat{c}^{\dagger}_{n s}(\mathbf{k}_{\alpha}) ,
    \\ \hat{b}_{\gamma \sigma} (\mathbf{k}_{\alpha}) &= \sum_{\{ns\}} \biggr( U_{\alpha}(\mathbf{k}_{\alpha})\biggr)_{\gamma\sigma,ns} \hat{c}_{n s}(\mathbf{k}_{\alpha}),
\end{align}
where the unitary $2 m_{\alpha} \times 2 m_{\alpha}$ matrix $U_{\alpha}(\mathbf{k}_{\alpha})$ satisfies the equation,
\begin{align}
    \biggr( U_{\alpha}^{\dagger}(\mathbf{k}_{\alpha}) \mathcal{H}_{\alpha}(\mathbf{k}_{\alpha}) U_{\alpha}(\mathbf{k}_{\alpha})\biggr)_{n s,n's'} = \epsilon_{\alpha,ns}(\mathbf{k}_{\alpha}) \delta_{ns,n's'},
\end{align}
with
\begin{align}
    \biggr(\mathcal{H}_{\alpha}(\mathbf{k}_{\alpha})\biggr)_{\gamma \sigma,\gamma'\sigma'} &= \sum_{\mathbf{R}_{\alpha\gamma}} t^{(\alpha)}_{\eta \eta'} (\mathbf{R}_{\alpha\gamma}- \mathbf{d}_{\alpha\gamma \gamma'})e^{i \mathbf{k}_{\alpha} \cdot (\mathbf{R}_{\alpha \gamma}-\mathbf{d}_{\alpha\gamma \gamma'})} - g_{\alpha}\mathbf{M}_{\alpha} \cdot \mathbf{S}_{\sigma \sigma'} \,\delta_{\gamma\gamma'}.
\end{align}

\section{Momentum Space Representation of Interlayer Hopping Hamiltonian}\label{Section Momentum Space Representation of Interlayer Hopping Hamiltonian}
Using the methodology outlined in Ref. \cite{Koshino2015}, we now consider the momentum space representation of the interlayer hopping Hamiltonian, which in the position space representation is given by
\begin{align}
    \hat{\mathcal{H}}_{\mathrm{int}} &= \sum_{\gamma,\gamma'}\sum_{\mathbf{R}_{1\gamma},\mathbf{R}_{2\gamma'}}  \sum_{\sigma,\sigma'}  t^{\mathrm{int}}_{\eta \eta'}(\mathbf{R}_{1\gamma}-\mathbf{R}_{2\gamma'}) \hat{b}^{\dagger}_{\sigma}(\mathbf{R}_{1\gamma})  \hat{b}_{\sigma'}(\mathbf{R}_{2\gamma'}) \nonumber + \mathrm{h.c.}
\end{align}
We emphasize that the layers are allowed to be translated by a vector $\mathbf{\Delta} = (\Delta_x,\Delta_y,\Delta_z)$ and rotated by an angle $\theta$ with respect to one another. Now, taking the Fourier transform of the interlayer hopping, we have
\begin{align}
   t_{\eta\eta'}^{\mathrm{int}}(\mathbf{R}_{1\gamma}-\mathbf{R}_{2\gamma'}) &= \int_{\mathbb{R}_2} \frac{d^2\mathbf{k}}{(2\pi)^2}\mathcal{T}_{\eta\eta'}^{\mathrm{int}}(\mathbf{k}) e^{i \mathbf{k}\cdot (\mathbf{R}_{1\gamma}-\mathbf{R}_{2\gamma'})},
\end{align}
where $\mathcal{T}_{\eta\eta'}^{\mathrm{int}}(\mathbf{k})$ is allowed to depend on $\Delta_z$. Using the momentum space respresentations of the operators $\hat{b}^{\dagger}_{\sigma}(\mathbf{R}_{1\gamma})$ and $\hat{b}_{\sigma}(\mathbf{R}_{2\gamma'})$, we can express the interlayer hopping Hamiltonian in the following form,
\begin{align}
    \hat{\mathcal{H}}_{\mathrm{int}} &= \sum_{\gamma, \gamma'} \sum_{\sigma, \sigma'} \sum_{\mathbf{k}_1,\mathbf{k}_2} \int_{\mathbb{R}_2} \frac{d^2 \mathbf{k}}{(2\pi)^2} \mathcal{T}_{\eta\eta'}^{\mathrm{int}}(\mathbf{k}) \,\hat{b}_{\gamma \sigma}^{\dagger}(\mathbf{k}_1) \hat{b}_{\gamma' \sigma'}(\mathbf{k}_2) \biggr(\sqrt{\frac{m_1}{N_1}} \sum_{\mathbf{R}_{1\gamma}}e^{i (\mathbf{k}+\mathbf{k}_1)\cdot \mathbf{R}_{1\gamma}}\biggr)\biggr(\sqrt{\frac{m_2}{N_2}} \sum_{\mathbf{R}_{2\gamma'}}e^{-i (\mathbf{k}+\mathbf{k}_2)\cdot \mathbf{R}_{2\gamma'}}\biggr) + \mathrm{h.c.}
\end{align}
We can now use the following well-known identities 
\begin{align}
    \frac{m_1}{N_1}\sum_{\mathbf{R}_{1\gamma}}e^{i (\mathbf{k}+\mathbf{k}_1)\cdot \mathbf{R}_{1\gamma}} &= \sum_{\mathbf{G}_1} \delta_{\mathbf{k}+\mathbf{k}_1,\mathbf{G}_1},  \\
     \frac{m_2}{N_2}\sum_{\mathbf{R}_{2\gamma'}}e^{-i (\mathbf{k}+\mathbf{k}_2)\cdot \mathbf{R}_{2\gamma'}} & = \sum_{\mathbf{G}_2} \delta_{\mathbf{k}+\mathbf{k}_2,\mathbf{G}_2} e^{-i \mathbf{G}_2 \cdot \boldsymbol{\Delta}}.
\end{align}
to arrive at
\begin{align}
    \hat{\mathcal{H}}_{\mathrm{int}} &= \sum_{\gamma, \gamma'} \sum_{\sigma, \sigma'} \sum_{\mathbf{k}_1,\mathbf{k}_2} \sum_{\mathbf{G}_1,\mathbf{G}_2} \sqrt{\frac{N_1 N_2}{m_1 m_2}}\biggr(\int_{\mathbb{R}_2} \frac{d^2 \mathbf{k}}{(2\pi)^2} \, \delta_{\mathbf{k}+\mathbf{k}_1,\mathbf{G}_1} \biggr) \mathcal{T}_{\eta\eta'}^{\mathrm{int}}(\mathbf{G}_1-\mathbf{k}_1)\,\delta_{\mathbf{k}_1-\mathbf{G}_1,\mathbf{k}_2-\mathbf{G}_2}\nonumber\\&\times e^{-i\mathbf{G}_2 \cdot \boldsymbol{\Delta}}\hat{b}_{\gamma \sigma}^{\dagger}(\mathbf{k}_1) \hat{b}_{\gamma' \sigma'}(\mathbf{k}_2)  + \mathrm{h.c.}
\end{align}
To evaluate the integral, we make use of the following relationship between the Kronecker delta and the Dirac delta function 
\begin{align}
   \lim_{A \rightarrow \infty} \delta_{\mathbf{k},\mathbf{k}'} &=\frac{(2\pi)^2}{A} \delta^{(2)}(\mathbf{k}-\mathbf{k}'),
\end{align}
where $A$ is the total area of a layer. Assuming both layers have equal areas, such that $A = \frac{N_1}{m_1} \mathcal{A}_1 = \frac{N_2}{m_2} \mathcal{A}_2$, with $\mathcal{A}_{\alpha}$ the area of a primitive unit cell of a sublattice in the layer $\alpha$, we find that
\begin{align}
    \hat{\mathcal{H}}_{\mathrm{int}} &= \sum_{\gamma, \gamma'} \sum_{\sigma, \sigma'} \sum_{\mathbf{k}_1,\mathbf{k}_2} \Gamma_{\eta\eta'}^{\mathrm{int}}(\mathbf{k}_1,\mathbf{k}_2) \hat{b}_{\gamma \sigma}^{\dagger}(\mathbf{k}_1) \hat{b}_{\gamma'\sigma'}(\mathbf{k}_2) + \, \mathrm{h.c.},
\end{align}
with
\begin{align}\label{Equation Gamma}
    \Gamma_{\eta\eta'}^{\mathrm{int}}(\mathbf{k}_1,\mathbf{k}_2) &= \frac{1}{\sqrt{\mathcal{A}_{1}\mathcal{A}_{2}}} \sum_{\mathbf{G}_1,\mathbf{G}_2} \mathcal{T}^{\mathrm{int}}_{\eta \eta'}(\mathbf{G}_1-\mathbf{k}_1) e^{-i\mathbf{G}_2 \cdot \mathbf{\Delta}} \delta_{\mathbf{k}_1-\mathbf{G}_1,\mathbf{k}_2-\mathbf{G}_2}.
\end{align}
Defining
\begin{align}
    \Upsilon_{ns,n's'}^{\mathrm{int}}(\mathbf{k}_1,\mathbf{k}_2) &= \sum_{\gamma,\gamma'} \sum_{\sigma,\sigma'} \biggr( U_1^{\dagger}(\mathbf{k}_1) \biggr)_{ns,\gamma\sigma} \Gamma_{\eta\eta'}^{\mathrm{int}}(\mathbf{k}_1,\mathbf{k}_2)\biggr( U_2(\mathbf{k}_2) \biggr)_{\gamma\sigma,ns}, 
\end{align}
we can also express $\hat{\mathcal{H}}_{\mathrm{int}}$ in terms of the $c$-operators,
\begin{align}
    \hat{\mathcal{H}}_{\mathrm{int}} &= \sum_{\{ns\}} \sum_{\{n's'\}} \sum_{\mathbf{k}_1,\mathbf{k}_2} \Upsilon_{ns,n's'}^{\mathrm{int}}(\mathbf{k}_1,\mathbf{k}_2) \hat{c}^{\dagger}_{ns}(\mathbf{k}_1) \hat{c}_{n's'}(\mathbf{k}_2) + \mathrm{h.c.}
\end{align}
Before moving on to calculate the perturbative corrections to the many-body ground state energy as a result of this interlayer interaction, we discuss an important property of the Fourier transform $\mathcal{T}_{\eta\eta'}^{\mathrm{int}}(\mathbf{k})$. For common types of parametrizations of $t^{\mathrm{int}}_{\eta \eta'}(\mathbf{r})$, such as the Slater-Koster parametrization \cite{Slater1954}, this Fourier transform will be a function that decays very rapidly for large $|\mathbf{k}|$. Consequently, one can usually restrict the summation over $\mathbf{G}_1$ and $\mathbf{G}_2$ in Eq.~\eqref{Equation Gamma} to only the first few shortest reciprocal lattice vectors, while still obtaining an accurate result \cite{Catarina2019}.

\section{First- and second-order perturbative corrections}

\subsection{Derivation of $E_0^{(1)}$ and $E_0^{(2)}$}
The unperturbed many-body ground state of the system is a filled Fermi sea given by
\begin{align}
    \ket{\Psi_0^0}&= \frac{1}{\mathcal{N}}\prod_{n,n',s}\prod_{\substack{ \mathbf{k}_1,\mathbf{k}_2\\ \epsilon_{1,ns}(\mathbf{k}_1) \leq \epsilon_{f} \\ \epsilon_{2,n' s}(\mathbf{k}_2) \leq \epsilon_f} } \hat{c}_{ns}^{\dagger}(\mathbf{k}_1) \hat{c}_{n's}^{\dagger}(\mathbf{k}_2) \ket{0},
\end{align}
where $\epsilon_f$ and $\ket{0}$ denote respectively the Fermi level and the vacuum state, and $\mathcal{N}$ is a normalization constant. From nondegenerate perturbation theory, we find that the corrections to the many-particle ground state energy are given, up to second order, by
\begin{align}
    E^{(1)}_0 &= \bra{\Psi_0^0} \hat{\mathcal{H}}_{\mathrm{int}} \ket{\Psi_0^0}, \\
    E^{(2)}_0 &= \sum_{m \neq 0} \frac{|\bra{\Psi_m^0} \hat{\mathcal{H}}_{\mathrm{int}} \ket{\Psi_0^{0}}|^2}{E_0^{(0)} - E_m^{(0)}}.
\end{align}
Here, $\ket{\Psi^0_m}$ is used to denote an excited many-particle eigenstate of the unperturbed Hamiltonian $\hat{\mathcal{H}}_1 + \hat{\mathcal{H}}_2$, with $E^{(0)}_m$ being the corresponding energy of said state. 

Since
\begin{align}
    \bra{\Psi_0^0} \hat{c}^{\dagger}_{ns}(\mathbf{k}_1) \hat{c}_{n's'}(\mathbf{k}_2) \ket{\Psi_0^0} &= 0, \\
    \bra{\Psi_0^0} \hat{c}^{\dagger}_{n's'}(\mathbf{k}_2) \hat{c}_{ns}(\mathbf{k}_1) \ket{\Psi_0^0} &= 0,
\end{align}
we immediately find that the first-order correction $E^{(1)}_0$ must vanish. Calculating the second-order correction to the ground state energy, $E^{(2)}_0$, is slightly more challenging. To reduce our computational effort, we first note that $\hat{\mathcal{H}}_{\mathrm{int}} \ket{\Psi_0^0}$ is equal to a superposition of states that differ from the many-particle ground state only by the presence of an electron-hole pair created from the Fermi sea. These pairs consist of either a hole in the Fermi sea of the first layer and an electron with energy $\epsilon_2 > \epsilon_f$ in the second layer or a hole in the Fermi sea of the second layer and an electron with energy $\epsilon_1 > \epsilon_f$ in the first layer. Thus, when calculating $E^{(2)}_0$, we need only consider excited states $\ket{\Psi_m^0}$ of the following form,
\begin{align}\label{Equation Excited States}
    \ket{\Psi_m^0} &= 
    \begin{cases}
        \hat{c}_{n s}^{\dagger}(\mathbf{k}_1) \hat{c}_{n's'}(\mathbf{k}_2) \ket{\Psi_0^0}. \\
        \hat{c}_{n's'}^{\dagger}(\mathbf{k}_2) \hat{c}_{n s}(\mathbf{k}_1) \ket{\Psi_0^0}. \\
    \end{cases}
\end{align}
Observing that the state $\hat{c}^{\dagger}_{n s}(\mathbf{k}_1) \hat{c}_{n' s'}(\mathbf{k}_2)\ket{\Psi_0^0}$ \,($\hat{c}^{\dagger}_{n's'}(\mathbf{k}_2) \hat{c}_{n s}(\mathbf{k}_1)\ket{\Psi_0^0}$) contains a hole in the Fermi sea of the second (first) layer and an extra electron in the first (second) layer, while the state $\bra{\Psi_0^0}\hat{c}^{\dagger}_{n s}(\mathbf{k}_1) \hat{c}_{n's'}(\mathbf{k}_2)$ ($\bra{\Psi_0^0} \hat{c}_{n's'}^{\dagger}(\mathbf{k}_2) \hat{c}_{n s}(\mathbf{k}_1)$) contains a hole in the Fermi sea of the first (second) layer and an extra electron in the second (first) layer, we find that
\begin{align}
    \bra{\Psi_0^0} \hat{c}_{\bar{n}\bar{s}}^{\dagger}(\mathbf{k}_1') \hat{c}_{\bar{n}'\bar{s}'}(\mathbf{k}_2')  \hat{c}^{\dagger}_{n s}(\mathbf{k}_1) \hat{c}_{n's'}(\mathbf{k}_2)\ket{\Psi_0^0} &= 0, \label{Equation Matrix Element Identity 1} \\
    \bra{\Psi_0^0} \hat{c}_{\bar{n}'\bar{s}'}^{\dagger}(\mathbf{k}_2') \hat{c}_{\bar{n}\bar{s}}(\mathbf{k}_1') \hat{c}^{\dagger}_{n's'}(\mathbf{k}_2) \hat{c}_{n s}(\mathbf{k}_1)  \ket{\Psi_0^0} &= 0.
\end{align}
On the other hand, using the anti-commutation relations $[ \hat{c}_{n s}(\mathbf{k}_1), \hat{c}^{\dagger}_{n's'}(\mathbf{k}_2) ]_+ = 0$ and $[ \hat{c}_{n s}(\mathbf{k}_1), \hat{c}_{n's'}(\mathbf{k}_2) ]_+ = 0$, we obtain 
\begin{align}
    \bra{\Psi_0^0} \hat{c}_{\bar{n}\bar{s}}^{\dagger}(\mathbf{k}_1') \hat{c}_{\bar{n}'\bar{s}'}(\mathbf{k}_2') \hat{c}^{\dagger}_{n's'}(\mathbf{k}_2) \hat{c}_{n s}(\mathbf{k}_1)  \ket{\Psi_0^0} &= \bra{\Psi_0^0} \hat{c}_{\bar{n}\bar{s}}^{\dagger}(\mathbf{k}_1') \hat{c}_{n s}(\mathbf{k}_1) \hat{c}_{\bar{n}'\bar{s}'}(\mathbf{k}_2') \hat{c}^{\dagger}_{n' s'}(\mathbf{k}_2)  \ket{\Psi_0^0}, \nonumber \\
    &= \delta_{\bar{n}n} \delta_{\bar{n}'n'} \delta_{\bar{s}s} \delta_{\bar{s}'s'} \delta_{\mathbf{k}_1',\mathbf{k}_1}\delta_{\mathbf{k}_2',\mathbf{k}_2} \,F(\epsilon_{1,ns}(\mathbf{k}_1))[1-F(\epsilon_{2,n's'}(\mathbf{k}_2))] \\
    \bra{\Psi_0^0} \hat{c}_{\bar{n}'\bar{s}'}^{\dagger}(\mathbf{k}_2') \hat{c}_{\bar{n}\bar{s}}(\mathbf{k}_1')  \hat{c}^{\dagger}_{n s}(\mathbf{k}_1) \hat{c}_{n's'}(\mathbf{k}_2)\ket{\Psi_0^0} &= \bra{\Psi_0^0}  \hat{c}_{\bar{n}\bar{s}}(\mathbf{k}_1')  \hat{c}^{\dagger}_{n s}(\mathbf{k}_1) \hat{c}_{\bar{n}'\bar{s}'}^{\dagger}(\mathbf{k}_2') \hat{c}_{n's'}(\mathbf{k}_2)\ket{\Psi_0^0} \nonumber \\
    &= \delta_{\bar{n}n} \delta_{\bar{n}'n'} \delta_{\bar{s}s} \delta_{\bar{s}'s'} \delta_{\mathbf{k}_1',\mathbf{k}_1}\delta_{\mathbf{k}_2',\mathbf{k}_2} \,F(\epsilon_{2,n's'}(\mathbf{k}_2))[1-F(\epsilon_{1,n s}(\mathbf{k}_1)] \label{Equation Matrix Element Identity 4}. 
\end{align}
Here, we have introduced the zero-temperature Fermi-Dirac distribution $F(\epsilon)$,
\begin{align}
    F(\epsilon) &= 1-H\big(\epsilon - \epsilon_f\big),
\end{align}
with $H(\epsilon)$ the Heaviside step function. For states of the form $\ket{\Psi_m^0} = \hat{c}^{\dagger}_{ns}(\mathbf{k}_1) \hat{c}_{n's'}(\mathbf{k}_2) \ket{\Psi_0^0}$, we then find that
\begin{align}
\big|\bra{\Psi_m^0}\hat{\mathcal{H}}_{\mathrm{int}}\ket{\Psi_0^0}\big|^2 &= |\Upsilon_{ns,n's'}^{\mathrm{int}}(\mathbf{k}_1,\mathbf{k}_2)|^2\,F(\epsilon_{2,n's'}(\mathbf{k}_2))[1-F(\epsilon_{1,n s}(\mathbf{k}_1)],
\end{align}
while for states of the form $\ket{\Psi_m^0} = \hat{c}^{\dagger}_{n's'}(\mathbf{k}_2) \hat{c}_{n s}(\mathbf{k}_1) \ket{\Psi_0^0}$ we find
\begin{align}
\big|\bra{\Psi_m^0}\hat{\mathcal{H}}_{\mathrm{int}}\ket{\Psi_0^0}\big|^2 &= |\Upsilon_{ns,n's'}^{\mathrm{int}}(\mathbf{k}_1,\mathbf{k}_2)|^2\,F(\epsilon_{1,n s}(\mathbf{k}_1))[1-F(\epsilon_{2,n' s'}(\mathbf{k}_2)].
\end{align}
Summing over all the relevant excited states $\ket{\Psi_m^0}$, we thus arrive at our final result,
\begin{align}\label{Equation Second-Order Correction}
     E_0^{(2)} = \sum_{\{ns\}} \sum_{\{n's'\}}\sum_{\mathbf{k}_1,\mathbf{k}_2} \bigg| \Upsilon^{\mathrm{int}}_{ns,n's'} (\mathbf{k}_1,\mathbf{k}_2) \bigg|^2 \biggr[\frac{F(\epsilon_{2,n's'}(\mathbf{k}_2))-F(\epsilon_{1,ns}(\mathbf{k}_1))}{\epsilon_{2,n's'}(\mathbf{k}_2)-\epsilon_{1,ns}(\mathbf{k}_1)}\biggr].
\end{align}

\subsection{Conversion to integral expression}
We now seek to convert Eq.~\eqref{Equation Second-Order Correction} into an integral expression. We start by introducing the following definitions,
\begin{align}
    \Gamma_{\eta \eta'}^{\mathrm{int}}(\mathbf{k}_1,\mathbf{k}_2,\mathbf{G}_1,\mathbf{G}_2) &= \frac{1}{\sqrt{\mathcal{A}_1\mathcal{A}_2}} \mathcal{T}_{\eta \eta'}(\mathbf{G}_1-\mathbf{k}_1) e^{-i\mathbf{G}_2 \cdot \boldsymbol{\Delta}},\\
    \Upsilon^{\mathrm{int}}_{ns,n's'}(\mathbf{k}_1,\mathbf{k}_2,\mathbf{G}_1,\mathbf{G}_2) &= \biggr(U_1^{\dagger}(\mathbf{k}_1)\,\Gamma^{\mathrm{int}}(\mathbf{k}_1,\mathbf{k}_2,\mathbf{G}_1,\mathbf{G}_2)\, U_2(\mathbf{k}_2)\biggr)_{ns,n's'},
\end{align}
such that
\begin{align}
    \Upsilon^{\mathrm{int}}_{ns,n's'}(\mathbf{k}_1,\mathbf{k}_2) &= \sum_{\mathbf{G}_1,\mathbf{G}_2} \Upsilon^{\mathrm{int}}_{ns,n's'}(\mathbf{k}_1,\mathbf{k}_2,\mathbf{G}_1,\mathbf{G}_2) \delta_{\mathbf{k}_1-\mathbf{G}_1,\mathbf{k}_2-\mathbf{G}_2}.
\end{align}
In the thermodynamic limit, we can make the replacement $\sum_{\mathbf{k}_{\alpha}} \rightarrow A\int_{\Omega_{\alpha}} d^2 \mathbf{k}_{\alpha}/(2\pi)^2 $, where $A$ is the total area of a single ferromagnetic layer and $\Omega_{\alpha}$ is the first Brillouin zone of the layer $\alpha$. Now, using the fact that
\begin{align}
    \delta_{\mathbf{k}_1-\mathbf{G}_1,\mathbf{k}_2-\mathbf{G}_2} \delta_{\mathbf{k}_1-\mathbf{G}_1',\mathbf{k}_2-\mathbf{G}_2'} = \delta_{\mathbf{k}_1-\mathbf{k}_2,\mathbf{G}_1-\mathbf{G}_2} \delta_{\mathbf{G}_{12},\mathbf{G}_{12}'},
\end{align}
where $\mathbf{G}_{12} = \mathbf{G}_1-\mathbf{G}_2$ and $\mathbf{G}_{12}' = \mathbf{G}_1'-\mathbf{G}_2'$, together with the identity
\begin{align}
    \lim_{A \rightarrow \infty} \delta_{\mathbf{k}_1-\mathbf{k}_2,\mathbf{G}_1-\mathbf{G}_2} = \frac{(2 \pi)^2}{A} \delta^{(2)}(\mathbf{k}_1-\mathbf{G}_{12}-\mathbf{k}_2),
\end{align}
we immediately find that
\begin{align}\label{Equation Second-Order Correction Integral Form}
    E_0^{(2)} &= \sum_{\{ns\}} \sum_{\{n's'\}} \sum_{\mathbf{G}_1,\mathbf{G}_2} \sum_{\mathbf{G_1',\mathbf{G}_2'}}A \,\delta_{\mathbf{G}_{12},\mathbf{G}_{12}'} \int_{\Omega_1} \frac{d^2 \mathbf{k}_1}{(2\pi)^2} \Phi_2(\mathbf{k}_1-\mathbf{G}_{12}) \Upsilon^{\mathrm{int}}_{ns,n's'}(\mathbf{k}_1,\mathbf{k}_1-\mathbf{G}_{12},\mathbf{G}_1,\mathbf{G}_2) \nonumber\\ &\times \biggr(\Upsilon^{\mathrm{int}}_{ns,n's'}(\mathbf{k}_1,\mathbf{k}_1-\mathbf{G}_{12},\mathbf{G}_1',\mathbf{G}_2')\biggr)^* \biggr[\frac{F(\epsilon_{2,n's'}(\mathbf{k}_1-\mathbf{G}_{12}))-F(\epsilon_{1,ns}(\mathbf{k}_1))}{\epsilon_{2,n's'}(\mathbf{k}_1-\mathbf{G}_{12})-\epsilon_{1,ns}(\mathbf{k}_1)}\biggr]. 
\end{align}
Here, we have defined the function $\Phi_2(\mathbf{k})$ as
\begin{align}
    \Phi_2(\mathbf{k}) &=
    \begin{cases}
        &1, \hspace{0.1cm} \text{if} \hspace{0.1cm} \mathbf{k} \in \Omega_2.\\
        &0, \hspace{0.1cm}\text{if} \hspace{0.1cm} \mathbf{k} \notin \Omega_2.
    \end{cases}
\end{align}

\subsection{Numerical evaluation procedure for the $E^{(2)}_0$-integral}

Although the integral of Eq.~\eqref{Equation Second-Order Correction} typically cannot be performed analytically, a numerical procedure to calculate its value can readily be given. However, because of the nonanalytic nature of the integrand, the intricacies of this integration procedure need to be carefully defined. 

We start by discussing how to deal with the zero-temperature Fermi-Dirac distribution $F(\epsilon)$ in a numerical integration setting. Observing that
\begin{align}
    F(\epsilon) = \lim_{T \rightarrow 0^+} F_{\scriptscriptstyle T}(\epsilon),
\end{align}
where we have defined $F_{\scriptscriptstyle T}(\epsilon)$ as
\begin{align}
    F_{\scriptscriptstyle T}(\epsilon) &= \frac{1}{1+e^{(\epsilon-\epsilon_f)/k_B T}},
\end{align}
we note that it is more convenient for numerical stability purposes to replace $F(\epsilon)$ in the integral by $F_{\scriptscriptstyle T}(\epsilon)$, with a small but nonzero value for the temperature $T$. This is because the integrand is proportional to a quotient of a Fermi function difference and a corresponding energy difference, which becomes problematic numerically when said energy difference is small. Working with a finite temperature $T$ allows us to deal with this problem, for a sufficiently small energy difference $\delta \epsilon$, by using the approximation
\begin{align}
    \frac{F_{\scriptscriptstyle T}(\epsilon_2)-F_{\scriptscriptstyle T}(\epsilon_1)}{\epsilon_2-\epsilon_1} \approx
        \frac{d F_{\scriptscriptstyle T}}{d \epsilon} \biggr|_{\epsilon_1}, \hspace{0.1cm} \text{if} \hspace{0.1cm} |\epsilon_2 -\epsilon_1| < \delta\epsilon,
\end{align}
where
\begin{align}\label{Equation Derivative Fermi-Dirac}
    \frac{d F_T(\epsilon)}{d \epsilon} &= \frac{e^{(\epsilon-\epsilon_f)/k_B T}}{k_B T(1-e^{(\epsilon-\epsilon_f)/k_B T})^2}.
\end{align}

Having appropriately dealt with the Fermi functions, we now discuss how to discretize the integration domain by introducing a suitable grid for the numerical integration procedure. While one might be tempted to use a uniform grid, we note that, in the limit of zero temperature, the right-hand side of Eq.\eqref{Equation Derivative Fermi-Dirac} becomes a Dirac delta function centered at the Fermi energy $\epsilon_f$. For small twist angles $\theta$ in particular, we expect the largest contribution to $E^{(2)}_0$ to come from the Fermi surface, and we thus discretize the first Brillouin zone with an adaptive grid to properly capture this contribution. We exclusively make use of SciPy's \textit{cubature} integration to accomplish this, although we note that for small $\theta$ more refined methods may be required. In this paper, we therefore mostly focus on twist angles $|\theta|>1^{\circ}$.

Finally, as already discussed in Section \ref{Section Momentum Space Representation of Interlayer Hopping Hamiltonian}, we will restrict the summation over the reciprocal lattice vectors to only the shortest few, thereby introducing an effective cut-off that still accurately approximates the exact result.

\section{Procedure for numerical extraction of interaction coefficients}
Here, we discuss how to extract the Heisenberg exchange constant $J(\theta)$, the Dzyaloshinskii-Moriya interaction (DMI) vector $\mathbf{D}(\theta)$ and the anisotropic exchange matrix $\mathcal{P}(\theta)$ from the second-order correction to the ground state energy, $E_0^{(2)}$. Now, assuming that $E_0^{(2)}$ is analytic in the exchange field, we can Taylor expand in terms of the vector components of $\mathbf{M}_1$ and $\mathbf{M}_2$ as follows,
\begin{align}\label{Equation Exchange Field Expansion}
    \frac{E_0^{(2)}}{A} &= C^{(00)}+\sum_{p \in \{x,y,z\}} \big(C^{(10)}_p  M_{1,p} + C^{(01)}_p  M_{2,p} \big)+\sum_{p,q \in \{x,y,z\}} \big( C_{p q}^{(11)}M_{1,p} M_{2,q} + C_{p q}^{(20)}M_{1,p} M_{1,q} +  C_{p q}^{(02)}M_{2,p} M_{2,q} \big) \nonumber \\&+ \text{higher-order terms in} \, \mathbf{M}_1 \, \text{and} \, \mathbf{M}_2.
\end{align}
We note that the coefficients $C_{pq}^{(11)}$ are directly related to $J$, $\mathbf{D}$, and $\mathcal{P}$ via the following equation,
\begin{align}
    \sum_{p,q \in \{x,y,z\}}  C_{p q}^{(11)}M_{1,p} M_{2,q} &= J \, \mathbf{M}_1 \cdot \mathbf{M}_2 + \mathbf{D} \cdot (\mathbf{M}_1 \times \mathbf{M}_2) + \mathbf{M}_1^T \, \mathcal{P} \, \mathbf{M}_2,
\end{align}
where the anisotropic exchange matrix $\mathcal{P}$ is both symmetric and traceless.

We discuss two approaches --- both based on the existence of this expansion --- to determine the Heisenberg exchange constant, the DMI vector and the anisotropic exchange matrix. The first method, which we refer to as the \textit{matrix method}, is computationally very efficient but more susceptible to numerical errors arising from the integration procedure. The second method, the \textit{method of least squares}, is more computationally demanding but significantly less sensitive to such errors. A detailed comparison between these methods is given in Section \ref{Section Numerical Results for the Bilayer Square Lattice System}, while in the main body of the paper we exclusively use the latter method.

\subsection{Matrix Method}
Since we are only interested in extracting the coefficients $C_{pq}^{(11)}$, we need to define a procedure that effectively filters out all the other $C$ coefficients, including those corresponding to the higher-order terms in $\mathbf{M}_1$ and $\mathbf{M}_2$. To do so, we first note that those terms containing an odd number of exchange field vector components \footnote{In general, we expect the coefficients corresponding to these terms to be zero anyway, although finding a proof of this conjecture is not trivial.} can be filtered out by taking the symmetric sum $E_0^{(2)}(\mathbf{M}_1,\mathbf{M}_2)+E_0^{(2)}(-\mathbf{M}_1,-\mathbf{M}_2)$. Furthermore, to distinguish the contributions from the $C_{pq}^{(11)}$-terms from those coming from the $C_{pq}^{(20)}$- and $C_{pq}^{(02)}$-terms, we can consider exchange field configurations of the form $(\mathbf{M}_1,\mathbf{0})$ and $(\mathbf{0},\mathbf{M}_2)$. Finally, to distinguish the $C_{pq}^{(11)}$-terms from the higher-order couplings between $\mathbf{M}_1$ and $\mathbf{M}_2$, we take $|\mathbf{M}_1|$ and $|\mathbf{M}_2|$ sufficiently small.

Let us now first determine the coefficients $C_{pq}^{(20)}$ and $C_{pq}^{(02)}$. For small enough $|\mathbf{M}_1|$ and $|\mathbf{M}_2|$, we expect that
\begin{align}
    \biggr|\sum_{k=3}^{\infty} \sum_{p_1,\cdots,p_k} \biggr[C_{p_1\cdots p_k}^{(k0)} \prod_{j=1}^k M_{1,p_j} + C_{p_1\cdots p_k}^{(0k)} \prod_{j=1}^k M_{2,p_j} \biggr] \biggr| \ll \max\big(|\mathbf{M}_1|^2,|\mathbf{M}_2|^2\big),
\end{align}
with $p_1,\cdots,p_k \in \{ x,y,z \}$, which allows us to safely ignore contributions to $E^{(2)}_0$ from terms higher than quadratic order in the exchange field. Introducing the abbreviation
\begin{align}
    E_S(\mathbf{M}_1,\mathbf{M}_2) = \frac{1}{2 A} \biggr( E_0^{(2)}(\mathbf{M}_1,\mathbf{M}_2)+E_0^{(2)}(-\mathbf{M}_1,-\mathbf{M}_2) \biggr),
\end{align}
we then obtain the following sets of linear equations for exchange field configurations of the form $(\mathbf{M}_1,\mathbf{0})$ or $(\mathbf{0},\mathbf{M}_2)$,
\begin{align}
    \begin{pmatrix}
        1 & 0 & 0 & 0 & 0 & 0 & 0 \\
        1 & 1 & 0 & 0 & 0 & 0 & 0 \\
        1 & 1/2 & 1 & 0 & 1/2 & 0 & 0 \\
        1 & 1/2 & 0 & 1 & 0 & 0 & 1/2 \\
        1 & 0 & 0 & 0 & 1 & 0 & 0 \\
        1 & 0 & 0 & 0 & 1/2 & 1 & 1/2 \\
        1 & 0 & 0 & 0 & 0 & 0 & 1
        \end{pmatrix}
\begin{pmatrix}
    C^{(00)} \\
    C^{(20)}_{xx} |\mathbf{M}_1|^2 \\
    C^{(20)}_{xy} |\mathbf{M}_1|^2 \\
    C^{(20)}_{xz} |\mathbf{M}_1|^2 \\
    C^{(20)}_{yy} |\mathbf{M}_1|^2 \\
    C^{(20)}_{yz} |\mathbf{M}_1|^2 \\
    C^{(20)}_{zz} |\mathbf{M}_1|^2
\end{pmatrix}
&=
\begin{pmatrix}
    E_S(\mathbf{0},\mathbf{0}) \\
    E_S(|\mathbf{M}_1|\hat{\mathbf{x}},\mathbf{0})  \\
    E_S(\frac{|\mathbf{M}_1|}{\sqrt{2}}\hat{\mathbf{x}}+\frac{|\mathbf{M}_1|}{\sqrt{2}}\hat{\mathbf{y}},\mathbf{0})  \\ E_S(\frac{|\mathbf{M}_1|}{\sqrt{2}}\hat{\mathbf{x}}+\frac{|\mathbf{M}_1|}{\sqrt{2}}\hat{\mathbf{z}},\mathbf{0}) \\
    E_S(|\mathbf{M}_1|\hat{\mathbf{y}},\mathbf{0}) \\
    E_S(\frac{|\mathbf{M}_1|}{\sqrt{2}}\hat{\mathbf{y}}+\frac{|\mathbf{M}_1|}{\sqrt{2}}\hat{\mathbf{z}},\mathbf{0}) \\
    E_S(|\mathbf{M}_1|\hat{\mathbf{z}},\mathbf{0})
\end{pmatrix}
,
\end{align}
and
\begin{align}
    \begin{pmatrix}
        1 & 0 & 0 & 0 & 0 & 0 & 0 \\
        1 & 1 & 0 & 0 & 0 & 0 & 0 \\
        1 & 1/2 & 1 & 0 & 1/2 & 0 & 0 \\
        1 & 1/2 & 0 & 1 & 0 & 0 & 1/2 \\
        1 & 0 & 0 & 0 & 1 & 0 & 0 \\
        1 & 0 & 0 & 0 & 1/2 & 1 & 1/2 \\
        1 & 0 & 0 & 0 & 0 & 0 & 1
        \end{pmatrix}
\begin{pmatrix}
    C^{(00)} \\
    C^{(02)}_{xx} |\mathbf{M}_2|^2 \\
    C^{(02)}_{xy} |\mathbf{M}_2|^2 \\
    C^{(02)}_{xz} |\mathbf{M}_2|^2\\
    C^{(02)}_{yy} |\mathbf{M}_2|^2\\
    C^{(02)}_{yz} |\mathbf{M}_2|^2\\
    C^{(02)}_{zz} |\mathbf{M}_2|^2
\end{pmatrix}
&=
\begin{pmatrix}
    E_S(\mathbf{0},\mathbf{0}) \\
    E_S(\mathbf{0},|\mathbf{M}_2|\hat{\mathbf{x}}) \\
    E_S(\mathbf{0},\frac{|\mathbf{M}_2|}{\sqrt{2}}\hat{\mathbf{x}}+\frac{|\mathbf{M}_2|}{\sqrt{2}}\hat{\mathbf{y}}) \\E_S(\mathbf{0},\frac{|\mathbf{M}_2|}{\sqrt{2}}\hat{\mathbf{x}}+\frac{|\mathbf{M}_2|}{\sqrt{2}}\hat{\mathbf{z}}) \\
    E_S(\mathbf{0},|\mathbf{M}_2|\hat{\mathbf{y}}) \\
    E_S(\mathbf{0},\frac{|\mathbf{M}_2|}{\sqrt{2}}\hat{\mathbf{y}}+\frac{|\mathbf{M}_2|}{\sqrt{2}}\hat{\mathbf{z}}) \\ 
    E_S(\mathbf{0},|\mathbf{M}_2|\hat{\mathbf{z}})
\end{pmatrix}
.
\end{align}
In deriving these equations, we have made use of the symmetry properties $C^{(20)}_{pq} = C^{(20)}_{qp}$ and $C^{(02)}_{pq} = C^{(02)}_{qp}$. We note that these equations are sufficient to determine all coefficients $C^{(20)}_{pq}$ and $C^{(02)}_{pq}$, as well as the constant $C^{(00)}$.

Having extracted the coefficients $C^{(20)}_{pq}$ and $C^{(02)}_{pq}$, we now turn our attention to the $J$, $\mathbf{D}$ and $\mathcal{P}$. For this purpose, we find it useful to introduce the following quantity,
\begin{align}
    \mathcal{E}_S(\mathbf{M}_1,\mathbf{M}_2) &= E_S(\mathbf{M}_1,\mathbf{M}_2) - C^{(00)}- \sum_{p,q \in \{x,y,z\}} \big( C_{p q}^{(20)}M_{1,p} M_{1,q} +  C_{p q}^{(02)}M_{2,p} M_{2,q} \big).
\end{align}
Again taking $|\mathbf{M}_1|$ and $|\mathbf{M}_2|$ to be sufficiently small, the procedure to extract the coefficients of interest is now given as follows.
\begin{enumerate}
    \item \textbf{Determining} $J$ \textbf{and the diagonal terms of} $\mathcal{P}$. For this, we consider a configuration where $\mathbf{M}_1 = \mathbf{M}_2$.
    Since $\Tr(\mathcal{P}) = 0$, we need only to consider three exchange field configurations to determine the four coefficients. We then obtain the following system of linear equations,
    \begin{align}
        \begin{cases}
            &\mathbf{M}_1 = \mathbf{M}_2 = |\mathbf{M}|\hat{\mathbf{x}} \hspace{0.3cm}\longrightarrow (J + \mathcal{P}_{xx})|\mathbf{M}|^2 = \mathcal{E}_S (|\mathbf{M}|\hat{\mathbf{x}}, |\mathbf{M}|\hat{\mathbf{x}}),\\
             &\mathbf{M}_1 = \mathbf{M}_2 = |\mathbf{M}|\hat{\mathbf{y}} \hspace{0.3cm}\longrightarrow (J + \mathcal{P}_{yy})|\mathbf{M}|^2 = \mathcal{E}_S (|\mathbf{M}|\hat{\mathbf{y}}, |\mathbf{M}|\hat{\mathbf{y}}),\\
              &\mathbf{M}_1 = \mathbf{M}_2 = |\mathbf{M}|\hat{\mathbf{z}}\hspace{0.3cm}\longrightarrow (J+ \mathcal{P}_{zz})|\mathbf{M}|^2 = \mathcal{E}_S (|\mathbf{M}|\hat{\mathbf{z}}, |\mathbf{M}|\hat{\mathbf{z}}).\\
        \end{cases}
    \end{align}
    
    \item \textbf{Determining} $\mathbf{D}$ \textbf{and the off-diagonal terms of} $\mathcal{P}$. For this, we consider a configuration where $\mathbf
    M_1 \perp \mathbf{M}_2$. We then obtain the following system of six linear equations,
    \begin{align}
        \begin{cases}
            &\mathbf{M}_1 = |\mathbf{M}|\hat{\mathbf{x}}, \, \mathbf{M}_2 = |\mathbf{M}|\hat{\mathbf{y}} \hspace{0.3cm}\longrightarrow (D_z + \mathcal{P}_{xy})|\mathbf{M}|^2 = \mathcal{E}_S (|\mathbf{M}|\hat{\mathbf{x}}, |\mathbf{M}|\hat{\mathbf{y}}),\\
             &\mathbf{M}_1 = |\mathbf{M}|\hat{\mathbf{y}}, \, \mathbf{M}_2 = |\mathbf{M}|\hat{\mathbf{x}} \hspace{0.3cm}\longrightarrow (-D_z + \mathcal{P}_{xy})|\mathbf{M}|^2 = \mathcal{E}_S (|\mathbf{M}|\hat{\mathbf{y}}, |\mathbf{M}|\hat{\mathbf{x}}),\\
              &\mathbf{M}_1 = |\mathbf{M}|\hat{\mathbf{y}}, \, \mathbf{M}_2 = |\mathbf{M}|\hat{\mathbf{z}} \hspace{0.3cm}\longrightarrow (D_x + \mathcal{P}_{yz})|\mathbf{M}|^2 = \mathcal{E}_S (|\mathbf{M}|\hat{\mathbf{y}}, |\mathbf{M}|\hat{\mathbf{z}}),\\
              &\mathbf{M}_1 = |\mathbf{M}|\hat{\mathbf{z}}, \, \mathbf{M}_2 = |\mathbf{M}|\hat{\mathbf{y}} \hspace{0.3cm}\longrightarrow (-D_x + \mathcal{P}_{yz})|\mathbf{M}|^2 = \mathcal{E}_S (|\mathbf{M}|\hat{\mathbf{z}}, |\mathbf{M}|\hat{\mathbf{y}}),\\
              &\mathbf{M}_1 = |\mathbf{M}|\hat{\mathbf{z}}, \, \mathbf{M}_2 = |\mathbf{M}|\hat{\mathbf{x}} \hspace{0.3cm}\longrightarrow (D_y + \mathcal{P}_{xz})|\mathbf{M}|^2 = \mathcal{E}_S (|\mathbf{M}|\hat{\mathbf{z}}, |\mathbf{M}|\hat{\mathbf{x}}),\\
              &\mathbf{M}_1 = |\mathbf{M}|\hat{\mathbf{x}}, \, \mathbf{M}_2 = |\mathbf{M}|\hat{\mathbf{z}} \hspace{0.3cm}\longrightarrow (-D_y + \mathcal{P}_{xz})|\mathbf{M}|^2 = \mathcal{E}_S (|\mathbf{M}|\hat{\mathbf{x}}, |\mathbf{M}|\hat{\mathbf{z}}),
        \end{cases}
    \end{align}
    which can be uniquely solved for the coefficients of interest.
    \end{enumerate}

\subsection{Method of Least Squares}

Here, instead of relying on a predetermined set of exchange field configurations, we consider a large amount of randomly generated configurations. The $C$-coefficients are then extracted by performing a global least-squares fit of the corresponding energies $E^{(2)}_0$ to the expansion of Eq.~\eqref{Equation Exchange Field Expansion} up to and including the second-order terms in the exchange fields.

Due to the large number of configurations considered, numerical integration errors are expected to average out, leading to more reliable estimates of $J$, $\mathbf{D}$ and $\mathcal{P}$. Note that we still take $|\mathbf{M}_1|$ and $|\mathbf{M}_2|$ to be sufficiently small, such that terms beyond second order in the exchange fields do not affect the fitting procedure.

\section{Calculations for the Bilayer Bravais System}\label{Section Calculations for the Bilayer Bravais System}

We now apply our theoretical framework to the simplest possible bilayer system that we can consider. We take both layers to be completely equivalent Bravais lattices, with the only possible differences between them being that $\mathbf{M}_1$ and $\mathbf{M}_2$ are allowed to point in different directions and that the layers are allowed to be rotated and translated with respect to one another. Furthermore, we consider intralayer hopping to be only between nearest neighbors, and we assume there is only one orbital state per lattice site. The interlayer hopping is assumed to be of the Slater-Koster form \cite{Slater1954},
\begin{align}\label{Equation Bravais Interlayer Hopping}
    t^{\mathrm{int}}_{\sigma \sigma'}(\mathbf{r}) = V^{\mathrm{int}}  e^{- r/r_0} \delta_{\sigma \sigma'}.
\end{align} 
Here, $r_0$ is the decay length of the Slater-Koster hopping and $V^{\mathrm{int}}$ is the interaction energy constant. Note that we have neglected interlayer spin-flip processes, and we have treated spin-up and spin-down electrons on equal footing. 

In what follows, we take $t^{\mathrm{int}}_{\sigma \sigma'}$ and $t^{(\alpha)}_{\sigma \sigma'}$ to correspond to the inter- and intralayer hopping coefficients as represented in the eigenbasis of the Pauli spin-$z$ operator, $\hat{\sigma}_z$. The eigenvectors of $\hat{\sigma}_z$ corresponding to the eigenvalues $+1$ and $-1$ are denoted by $\ket{\uparrow}$ and $\ket{\downarrow}$, respectively.

\subsection{Consequences of rotational symmetries}
Since we only consider nearest-neighbor intralayer hopping and we are only dealing with one sublattice per layer, we find that the intralayer tight-binding coefficients are of the following form,
\begin{align}
    t_{\sigma\sigma'}^{(\alpha)}(\mathbf{R}_{\alpha}-\mathbf{R}_{\alpha}') &= \sum_{\boldsymbol{\tau}_{\alpha}} t_{\sigma\sigma'}^{(\alpha)}(\boldsymbol{\tau}_{\alpha}) \, \delta_{\mathbf{R}_{\alpha}-\mathbf{R}_{\alpha}',\boldsymbol{\tau}_{\alpha}},
\end{align}
where we sum over all nearest-neighbor vectors $\boldsymbol{\tau}_{\alpha}$. By considering the discrete rotational symmetries of the lattice, we can find the relationship between the tight-binding coefficients $t_{\sigma\sigma'}^{(\alpha)}(\boldsymbol{\tau}_{\alpha})$ and $t_{\sigma\sigma'}^{(\alpha)}(\boldsymbol{\tau}_{\alpha}')$, with $\boldsymbol{\tau}_{\alpha} \neq \boldsymbol{\tau}_{\alpha}'$. We note that these relationships are required in order to derive $\mathcal{H}_{\alpha}(\mathbf{k})$ for different types of two-dimensional Bravais lattices.

In what follows, we will find it useful to consider the single-particle tight-binding Hamiltonian $\hat{H}_{\alpha,\mathrm{tb}}$ of the layer $\alpha$ in first-quantized form,
\begin{equation}
    \hat{H}_{\alpha,\mathrm{tb}} = \sum_{\mathbf{R}_{\alpha}} \sum_{\boldsymbol{\tau}_{\alpha}} \sum_{\sigma,\sigma'} t_{\sigma \sigma'}^{(\alpha)}(\boldsymbol{\tau}_{\alpha}) \ket{\mathbf{R}_{\alpha}, \sigma} \bra{\mathbf{R}_{\alpha}-\boldsymbol{\tau}_{\alpha}, \sigma'},
\end{equation}
where the Hermiticity of $\hat{H}_{\alpha,\mathrm{tb}}$ requires that
\begin{align}\label{Equation Hermiticity Condition}
    t_{\sigma' \sigma}^{(\alpha)}(-\boldsymbol{\tau}_{\alpha}) &= \big( t_{\sigma \sigma'}^{(\alpha)}(\boldsymbol{\tau}_{\alpha})\big)^*.
\end{align}
We note that $\hat{H}_{\alpha,\mathrm{tb}}$ is related to its second-quantized form $\hat{\mathcal{H}}_{\alpha,\mathrm{tb}}$ in the standard way,
\begin{align}
    \hat{\mathcal{H}}_{\alpha,\mathrm{tb}} &= \sum_{\bar{\mathbf{R}}_{\alpha},\bar{\mathbf{R}}_{\alpha}'} \sum_{\bar{\sigma},\bar{\sigma}'} \bra{\bar{\mathbf{R}}_{\alpha}, \bar{\sigma}} \hat{H}_{\alpha,\mathrm{tb}} \ket{\bar{\mathbf{R}}_{\alpha}', \bar{\sigma}'} \hat{\alpha}^{\dagger}_{\bar{\sigma}}(\bar{\mathbf{R}}_{\alpha}) \hat{\alpha}_{\bar{\sigma}'}(\bar{\mathbf{R}}_{\alpha}').
\end{align}

Denoting by $\mathcal{M}(\phi)$ the $3 \times 3$ matrix which rotates a position vector $\mathbf{r}$ around the $z$-axis by an angle $\phi$, we first wish to find its representation $\hat{\mathcal{D}}(\phi) \equiv \hat{\mathcal{D}}(\mathcal{M}(\phi))$ on the Hilbert space of the single-layer ferromagnet. This representation is given by
\begin{equation}
    \hat{\mathcal{D}}(\phi) = e^{- i \phi \hat{L}_z/\hbar} \otimes e^{- i \phi \hat{\sigma}_z/2},
\end{equation}
with $\hat{L}_z$ the operator corresponding to the $z$-component of the orbital angular momentum. Now, due to the lattice symmetries, there exists a discrete set of angles $\phi_n$ for which the rotation operator $\hat{\mathcal{D}}(\phi_n)$ will be a symmetry operation, in the sense that
\begin{equation}\label{Equation Symmetry Condition}
    \hat{\mathcal{D}}^{\dagger}(\phi_n) \, \hat{H}_{\alpha,\mathrm{tb}}\,\hat{\mathcal{D}}(\phi_n) = \hat{H}_{\alpha,\mathrm{tb}}.
\end{equation}
Using Eq.~\eqref{Equation Symmetry Condition} together with the following identities,
\begin{align}
    \begin{cases}
    \bra{\mathbf{R}_{\alpha}, \uparrow} \hat{\mathcal{D}}^{\dagger}(\phi_n) \, \hat{H}_{\alpha,\mathrm{tb}}\,\hat{\mathcal{D}}(\phi_n) \ket{\mathbf{R}_{\alpha}', \uparrow} &= \bra{\mathcal{M}(\phi_n)\mathbf{R}_{\alpha}, \uparrow} \hat{H}_{\alpha,\mathrm{tb}} \ket{\mathcal{M}(\phi_n)\mathbf{R}_{\alpha}', \uparrow},   \\
    \bra{\mathbf{R}_{\alpha}, \uparrow} \hat{\mathcal{D}}^{\dagger}(\phi_n) \, \hat{H}_{\alpha,\mathrm{tb}}\,\hat{\mathcal{D}}(\phi_n) \ket{\mathbf{R}_{\alpha}', \downarrow} &=  \bra{\mathcal{M}(\phi_n)\mathbf{R}_{\alpha}, \uparrow} \hat{H}_{\alpha,\mathrm{tb}} \ket{\mathcal{M}(\phi_n)\mathbf{R}_{\alpha}', \downarrow} e^{i \phi_n},\\
    \bra{\mathbf{R}_{\alpha}, \downarrow} \hat{\mathcal{D}}^{\dagger}(\phi_n) \, \hat{H}_{\alpha,\mathrm{tb}}\,\hat{\mathcal{D}}(\phi_n) \ket{\mathbf{R}_{\alpha}', \uparrow} &= \bra{\mathcal{M}(\phi_n)\mathbf{R}_{\alpha}, \downarrow} \hat{H}_{\alpha,\mathrm{tb}} \ket{\mathcal{M}(\phi_n)\mathbf{R}_{\alpha}', \uparrow} e^{-i \phi_n}, \\
    \bra{\mathbf{R}_{\alpha}, \downarrow} \hat{\mathcal{D}}^{\dagger}(\phi_n) \, \hat{H}_{\alpha,\mathrm{tb}}\,\hat{\mathcal{D}}(\phi_n) \ket{\mathbf{R}_{\alpha}', \downarrow} &= \bra{\mathcal{M}(\phi_n)\mathbf{R}_{\alpha}, \downarrow} \hat{H}_{\alpha,\mathrm{tb}} \ket{\mathcal{M}(\phi_n)\mathbf{R}_{\alpha}', \downarrow},
    \end{cases}
\end{align}
we immediately find that
\begin{equation}
    \bra{\mathbf{R}_{\alpha}, \sigma} \hat{H}_{\alpha,\mathrm{tb}} \ket{\mathbf{R}_{\alpha}', \sigma'} =  \bra{\mathcal{M}(\phi_n) \mathbf{R}_{\alpha},  \sigma} \hat{H}_{\alpha,\mathrm{tb}} \ket{\mathcal{M}(\phi_n) \mathbf{R}_{\alpha}',  \sigma'} e^{i \, \mathcal{G}(\sigma, \sigma') \phi_n}.
\end{equation}
Here, we have introduced the following definition,
\begin{align}
    \mathcal{G}(\sigma,\sigma') =
    \begin{cases}
         1, \hspace{0.15cm} &\text{for} \hspace{0.15cm} \sigma = \uparrow, \sigma' = \downarrow. \\
        -1, \hspace{0.15cm} &\text{for} \hspace{0.15cm} \sigma = \downarrow, \sigma' = \uparrow. \\
         0, \hspace{0.15cm} &\text{else}.
    \end{cases}
\end{align}
We thus find that hopping coefficients corresponding to different directions are connected by rotational symmetries via the following equation,
\begin{align}
    t_{\sigma \sigma'}^{(\alpha)}(\boldsymbol{\tau}_{\alpha}) = t_{\sigma \sigma'}^{(\alpha)}(\mathcal{M}(\phi_n) \boldsymbol{\tau}_{\alpha}) e^{i \, \mathcal{G}(\sigma, \sigma') \phi_n}. \label{Equation Rotational Symmetry Condition}
\end{align}
Incidentally, we note that a combination of Eq.~\eqref{Equation Rotational Symmetry Condition} with Eq.~\eqref{Equation Hermiticity Condition} immediately yields a condition on the spin-flip same-direction hopping coefficients as well, namely
\begin{align}\label{Equation Spin-Flip Hopping Condition}
    t_{\downarrow \uparrow}^{(\alpha)}(\boldsymbol{\tau}_{\alpha}) = -\big(t_{\uparrow \downarrow}^{(\alpha)}(\boldsymbol{\tau}_{\alpha})\big)^*.
\end{align}

\subsection{Relevant properties of all 2D Bravais lattices}
In two dimensions, there are exactly five Bravais lattice structures: square, rectangular, centered rectangular, oblique and triangular. In this section, we will provide all their properties that are needed to calculate the second-order correction to the ground-state energy. These properties include the primitive lattice basis vectors and the corresponding reciprocal lattice basis vectors, as well as their symmetry angles $\phi_n$ and the resulting expression for $\hat{\mathcal{H}}_{\alpha}(\mathbf{k}_{\alpha})$.

The primitive lattice vectors are denoted by $\mathbf{a}_{1,\alpha}$ and $\mathbf{a}_{2,\alpha}$, with the lattice sites $\mathbf{R}_{\alpha}$ in the layer $\alpha$ then being given by
\begin{align}
    \begin{cases}
    \mathbf{R}_{1} &= n_{1,1} \, \mathbf{a}_{1,1} \,+ n_{2,1} \, \mathbf{a}_{2,1},\\
    \mathbf{R}_{2} &= n_{1,2} \, \mathbf{a}_{1,2} \,+ n_{2,2} \, \mathbf{a}_{2,2} + \boldsymbol{\Delta},
    \end{cases}
\end{align}
where $n_{1,\alpha},n_{2,\alpha} \in \mathbb{Z}$. The corresponding reciprocal lattice basis vectors are denoted by $\mathbf{b}_{1,\alpha}$ and $\mathbf{b}_{2,\alpha}$, with any point $\mathbf{G}_{\alpha}$ in the reciprocal lattice of layer $\alpha$ then being given by
\begin{align}
    \mathbf{G}_{\alpha} = \ell_{1,\alpha} \, \mathbf{b}_{1,\alpha} \,+ \ell_{2,\alpha} \, \mathbf{b}_{2,\alpha},
\end{align}
where $\ell_{1,\alpha},\ell_{2,\alpha} \in \mathbb{Z}$. Finally, to take into account the fact that the layers are typically rotated with respect to one another by an angle $\theta$, we introduce Cartesian vector pairs $\{\hat{\mathbf{x}}_1,\hat{\mathbf{y}}_1\}$ and $\{\hat{\mathbf{x}}_2,\hat{\mathbf{y}}_2\}$, which are related to one another via the following equations,
\begin{align}
    \begin{cases}
    \hat{\mathbf{x}}_2 &= \cos(\theta) \, \hat{\mathbf{x}}_1 + \sin(\theta) \, \hat{\mathbf{y}}_1,\\
    \hat{\mathbf{y}}_2 &= -\sin(\theta) \, \hat{\mathbf{x}}_1 + \cos(\theta) \, \hat{\mathbf{y}}_1.
    \end{cases}
\end{align}

\subsubsection{Square lattice}
For the square lattice, we use the following pair of primitive basis vectors,
\begin{align}
    \begin{cases}
    \mathbf{a}_{1,\alpha} &= a \, \hat{\mathbf{x}}_{\alpha},\\
    \mathbf{a}_{2,\alpha} &= a \, \hat{\mathbf{y}}_{\alpha}.
    \end{cases}
\end{align}
The corresponding reciprocal lattice basis vectors are given by
\begin{align}
    \begin{cases}
    \mathbf{b}_{1,\alpha} &= \frac{2\pi}{a} \, \hat{\mathbf{x}}_{\alpha},\\
    \mathbf{b}_{2,\alpha} &= \frac{2\pi}{a} \, \hat{\mathbf{y}}_{\alpha}.
    \end{cases}
\end{align}
The symmetry angles $\phi_n$ are the elements of the set $\{0,\frac{\pi}{2},\pi,\frac{3\pi}{2}\}$. Introducing the abbreviation $t_{\sigma \sigma'}^{(\alpha)}(a \hat{\mathbf{x}}_{\alpha}) = t_{\sigma \sigma'}^{(\alpha)}$, and applying Eqs.~\eqref{Equation Rotational Symmetry Condition} and \eqref{Equation Spin-Flip Hopping Condition}, we immediately find that
\begin{align}
    \begin{cases}
    \bigg(\mathcal{H}_{\alpha}(\mathbf{k}_{\alpha})\bigg)_{\uparrow\uparrow} &= 2\, t^{(\alpha)}_{\uparrow \uparrow} \biggr[ \cos(k_{x_{\alpha}} a) + \cos(k_{y_{\alpha}} a)\biggr] - g_{\alpha} \, M_{\alpha,z}, \\
    \bigg(\mathcal{H}_{\alpha}(\mathbf{k}_{\alpha})\bigg)_{\uparrow\downarrow} &= 2\, t^{(\alpha)}_{\uparrow \downarrow} \biggr[ \sin(k_{y_{\alpha}} a) + i \sin(k_{x_{\alpha}} a)\biggr] - g_{\alpha}\big(M_{\alpha,x}-i M_{\alpha,y}\big),\\
    \bigg(\mathcal{H}_{\alpha}(\mathbf{k}_{\alpha})\bigg)_{\downarrow\uparrow} &= 2\, \big(t^{(\alpha)}_{\uparrow \downarrow}\big)^* \biggr[ \sin(k_{y_{\alpha}} a) - i \sin(k_{x_{\alpha}} a)\biggr] - g_{\alpha}\big(M_{\alpha,x}+i M_{\alpha,y}\big),\\
    \bigg(\mathcal{H}_{\alpha}(\mathbf{k}_{\alpha})\bigg)_{\downarrow\downarrow} &= 2\, t^{(\alpha)}_{\downarrow \downarrow} \biggr[ \cos(k_{x_{\alpha}} a) + \cos(k_{y_{\alpha}} a)\biggr] + g_{\alpha} \, M_{\alpha,z}.
    \end{cases}
\end{align}

\subsubsection{Rectangular lattice}
For the rectangular lattice, we use the following pair of primitive basis vectors,
\begin{align}
    \begin{cases}
    \mathbf{a}_{1,\alpha} &= a \, \hat{\mathbf{x}}_{\alpha},\\
    \mathbf{a}_{2,\alpha} &= b \, \hat{\mathbf{y}}_{\alpha}.
    \end{cases}
\end{align}
The corresponding reciprocal lattice basis vectors are given by
\begin{align}
    \begin{cases}
    \mathbf{b}_{1,\alpha} &= \frac{2\pi}{a} \, \hat{\mathbf{x}}_{\alpha},\\
    \mathbf{b}_{2,\alpha} &= \frac{2\pi}{b} \, \hat{\mathbf{y}}_{\alpha}.
    \end{cases}
\end{align}
The symmetry angles $\phi_n$ are the elements of the set $\{0,\pi\}$. Introducing the abbreviations $t_{\sigma \sigma'}^{(\alpha)}(a \hat{\mathbf{x}}_{\alpha}) = t_{\sigma \sigma'}^{(\alpha,x)}$ and $t_{\sigma \sigma'}^{(\alpha)}(b \hat{\mathbf{y}}_{\alpha}) = t_{\sigma \sigma'}^{(\alpha,y)}$, and applying Eqs.~\eqref{Equation Rotational Symmetry Condition} and \eqref{Equation Spin-Flip Hopping Condition}, we immediately find that
\begin{align}
    \begin{cases}
    \bigg(\mathcal{H}_{\alpha}(\mathbf{k}_{\alpha})\bigg)_{\uparrow\uparrow} &= 2\, t^{(\alpha,x)}_{\uparrow \uparrow} \cos(k_{x_{\alpha}} a) + 2\, t^{(\alpha,y)}_{\uparrow \uparrow}\cos(k_{y_{\alpha}} b) - g_{\alpha} \, M_{\alpha,z}, \\
    \bigg(\mathcal{H}_{\alpha}(\mathbf{k}_{\alpha})\bigg)_{\uparrow\downarrow} &= 2i\, t^{(\alpha,x)}_{\uparrow \downarrow} \sin(k_{x_{\alpha}} a) + 2i\, t^{(\alpha,y)}_{\uparrow \downarrow} \sin(k_{y_{\alpha}} b) - g_{\alpha}\big(M_{\alpha,x}-i M_{\alpha,y}\big),\\
    \bigg(\mathcal{H}_{\alpha}(\mathbf{k}_{\alpha})\bigg)_{\downarrow\uparrow} &= -2i\, \big(t^{(\alpha,x)}_{\uparrow \downarrow}\big)^* \sin(k_{x_{\alpha}} a) - 2i\, \big(t^{(\alpha,y)}_{\uparrow \downarrow})^* \sin(k_{y_{\alpha}} b)  - g_{\alpha}\big(M_{\alpha,x}+i M_{\alpha,y}\big),\\
    \bigg(\mathcal{H}_{\alpha}(\mathbf{k}_{\alpha})\bigg)_{\downarrow\downarrow} &= 2\, t^{(\alpha,x)}_{\downarrow \downarrow} \cos(k_{x_{\alpha}} a) + 2\, t^{(\alpha,y)}_{\downarrow \downarrow}\cos(k_{y_{\alpha}} b) + g_{\alpha} \, M_{\alpha,z}.
    \end{cases}
\end{align}

\subsubsection{Centered rectangular lattice}
For the centered rectangular lattice, we use the following pair of primitive basis vectors,
\begin{align}
    \begin{cases}
    \mathbf{a}_{1,\alpha} &= \frac{a}{2} \, \hat{\mathbf{x}}_{\alpha}+\frac{b}{2} \hat{\mathbf{y}}_{\alpha},\\
    \mathbf{a}_{2,\alpha} &= \frac{a}{2} \, \hat{\mathbf{x}}_{\alpha}-\frac{b}{2} \hat{\mathbf{y}}_{\alpha}.
    \end{cases}
\end{align}
The corresponding reciprocal lattice basis vectors are given by
\begin{align}
    \begin{cases}
    \mathbf{b}_{1,\alpha} &= \frac{2\pi}{a} \, \hat{\mathbf{x}}_{\alpha}+\frac{2\pi}{b} \hat{\mathbf{y}}_{\alpha},\\
    \mathbf{b}_{2,\alpha} &= \frac{2\pi}{a} \, \hat{\mathbf{x}}_{\alpha}-\frac{2\pi}{b} \hat{\mathbf{y}}_{\alpha}.
    \end{cases}
\end{align}
The symmetry angles $\phi_n$ are the elements of the set $\{0,\pi\}$.  Introducing the abbreviations $t_{\sigma \sigma'}^{(\alpha)}(\frac{a}{2} \hat{\mathbf{x}}_{\alpha}+\frac{b}{2} \hat{\mathbf{y}}_{\alpha}) = t_{\sigma \sigma'}^{(\alpha,+)}$ and $t_{\sigma \sigma'}^{(\alpha)}(\frac{a}{2} \hat{\mathbf{x}}_{\alpha}-\frac{b}{2} \hat{\mathbf{y}}_{\alpha}) = t_{\sigma \sigma'}^{(\alpha,-)}$, and applying Eqs.~\eqref{Equation Rotational Symmetry Condition} and \eqref{Equation Spin-Flip Hopping Condition}, we immediately find that
\begin{align}
    \begin{cases}
    \bigg(\mathcal{H}_{\alpha}(\mathbf{k}_{\alpha})\bigg)_{\uparrow\uparrow} &= 2\, t^{(\alpha,+)}_{\uparrow \uparrow} \cos(\frac{1}{2} k_{x_{\alpha}} a+\frac{1}{2} k_{y_{\alpha}} b) + 2\, t^{(\alpha,-)}_{\uparrow \uparrow}\cos(\frac{1}{2} k_{x_{\alpha}} a-\frac{1}{2} k_{y_{\alpha}} b) - g_{\alpha} \, M_{\alpha,z}, \\
    \bigg(\mathcal{H}_{\alpha}(\mathbf{k}_{\alpha})\bigg)_{\uparrow\downarrow} &= 2i\, t^{(\alpha,+)}_{\uparrow \downarrow} \sin(\frac{1}{2} k_{x_{\alpha}} a+\frac{1}{2} k_{y_{\alpha}} b) + 2i\, t^{(\alpha,-)}_{\uparrow \downarrow} \sin(\frac{1}{2} k_{x_{\alpha}} a-\frac{1}{2} k_{y_{\alpha}} b) - g_{\alpha}\big(M_{\alpha,x}-i M_{\alpha,y}\big),\\
    \bigg(\mathcal{H}_{\alpha}(\mathbf{k}_{\alpha})\bigg)_{\downarrow\uparrow} &= -2i\, \big(t^{(\alpha,+)}_{\uparrow \downarrow}\big)^* \sin(\frac{1}{2} k_{x_{\alpha}} a+\frac{1}{2} k_{y_{\alpha}} b) - 2i\, \big(t^{(\alpha,-)}_{\uparrow \downarrow})^* \sin(\frac{1}{2} k_{x_{\alpha}} a-\frac{1}{2} k_{y_{\alpha}} b)  - g_{\alpha}\big(M_{\alpha,x}+i M_{\alpha,y}\big),\\
    \bigg(\mathcal{H}_{\alpha}(\mathbf{k}_{\alpha})\bigg)_{\downarrow\downarrow} &= 2\, t^{(\alpha,+)}_{\downarrow \downarrow} \cos(\frac{1}{2} k_{x_{\alpha}} a+\frac{1}{2} k_{y_{\alpha}} b) + 2\, t^{(\alpha,-)}_{\downarrow \downarrow}\cos(\frac{1}{2} k_{x_{\alpha}} a-\frac{1}{2} k_{y_{\alpha}} b) + g_{\alpha} \, M_{\alpha,z}.
    \end{cases}
\end{align}

\subsubsection{Oblique lattice}
For the oblique lattice, we use the following pair of primitive basis vectors,
\begin{align}
    \begin{cases}
    \mathbf{a}_{1,\alpha} &= a \, \hat{\mathbf{x}}_{\alpha},\\
    \mathbf{a}_{2,\alpha} &= b \, \hat{\mathbf{x}}_{\alpha}+c \, \hat{\mathbf{y}}_{\alpha}.
    \end{cases}
\end{align}
The corresponding reciprocal lattice basis vectors are given by
\begin{align}
    \begin{cases}
    \mathbf{b}_{1,\alpha} &= \frac{2\pi}{a} \, \hat{\mathbf{x}}_{\alpha}-\frac{2\pi \,b}{a c} \, \hat{\mathbf{y}}_{\alpha},\\
    \mathbf{b}_{2,\alpha} &= \frac{2\pi}{c} \, \hat{\mathbf{y}}_{\alpha}.
    \end{cases}
\end{align}
The symmetry angles $\phi_n$ are the elements of the set $\{0,\pi\}$. Introducing the abbreviations $t_{\sigma \sigma'}^{(\alpha)}(a \hat{\mathbf{x}}_{\alpha}) = t_{\sigma \sigma'}^{(\alpha,x)}$ and $t_{\sigma \sigma'}^{(\alpha)}(b \hat{\mathbf{x}}_{\alpha}+c \hat{\mathbf{y}}_{\alpha}) = t_{\sigma \sigma'}^{(\alpha,xy)}$, and applying Eqs.~\eqref{Equation Rotational Symmetry Condition} and \eqref{Equation Spin-Flip Hopping Condition}, we immediately find that
\begin{align}
    \begin{cases}
    \bigg(\mathcal{H}_{\alpha}(\mathbf{k}_{\alpha})\bigg)_{\uparrow\uparrow} &= 2\, t^{(\alpha,x)}_{\uparrow \uparrow} \cos(k_{x_{\alpha}} a) + 2\, t^{(\alpha,xy)}_{\uparrow \uparrow}\cos(k_{x_{\alpha}} b+k_{y_{\alpha}} c) - g_{\alpha} \, M_{\alpha,z}, \\
    \bigg(\mathcal{H}_{\alpha}(\mathbf{k}_{\alpha})\bigg)_{\uparrow\downarrow} &= 2i\, t^{(\alpha,x)}_{\uparrow \downarrow} \sin(k_{x_{\alpha}} a) + 2i\, t^{(\alpha,xy)}_{\uparrow \downarrow} \sin(k_{x_{\alpha}} b+k_{y_{\alpha}} c) - g_{\alpha}\big(M_{\alpha,x}-i M_{\alpha,y}\big),\\
    \bigg(\mathcal{H}_{\alpha}(\mathbf{k}_{\alpha})\bigg)_{\downarrow\uparrow} &= -2i\, \big(t^{(\alpha,x)}_{\uparrow \downarrow}\big)^* \sin(k_{x_{\alpha}} a) - 2i\, \big(t^{(\alpha,xy)}_{\uparrow \downarrow})^* \sin(k_{x_{\alpha}} b+k_{y_{\alpha}} c)  - g_{\alpha}\big(M_{\alpha,x}+i M_{\alpha,y}\big),\\
    \bigg(\mathcal{H}_{\alpha}(\mathbf{k}_{\alpha})\bigg)_{\downarrow\downarrow} &= 2\, t^{(\alpha,x)}_{\downarrow \downarrow} \cos(k_{x_{\alpha}} a) + 2\, t^{(\alpha,xy)}_{\downarrow \downarrow}\cos(k_{x_{\alpha}} b+k_{y_{\alpha}} c) + g_{\alpha} \, M_{\alpha,z}.
    \end{cases}
\end{align}

\subsubsection{Triangular lattice}
For the triangular lattice, we use the following pair of primitive basis vectors,
\begin{align}
    \begin{cases}
    \mathbf{a}_{1,\alpha} &= a \, \hat{\mathbf{x}}_{\alpha},\\
    \mathbf{a}_{2,\alpha} &= \frac{a}{2} \, \hat{\mathbf{x}}_{\alpha} + \frac{\sqrt{3}a}{2} \, \hat{\mathbf{y}}_{\alpha}.
    \end{cases}
\end{align}
The corresponding reciprocal lattice basis vectors are given by
\begin{align}
    \begin{cases}
    \mathbf{b}_{1,\alpha} &= \frac{2\pi}{a} \, \hat{\mathbf{x}}_{\alpha} - \frac{2\pi}{\sqrt{3}a} \, \hat{\mathbf{y}}_{\alpha},\\
    \mathbf{b}_{2,\alpha} &= \frac{4\pi}{\sqrt{3}a} \, \hat{\mathbf{y}}_{\alpha}.
    \end{cases}
\end{align}
The symmetry angles $\phi_n$ are the elements of the set $\{0,\frac{\pi}{3},\frac{2\pi}{3}\pi,\frac{4\pi}{3},\frac{5\pi}{3}\}$. Introducing the abbreviation $t_{\sigma \sigma'}^{(\alpha)}(a \hat{\mathbf{x}}_{\alpha}) = t_{\sigma \sigma'}^{(\alpha)}$, and applying Eqs.~\eqref{Equation Rotational Symmetry Condition} and \eqref{Equation Spin-Flip Hopping Condition}, we immediately find that

\begin{align}
    \begin{cases}
    \bigg(\mathcal{H}_{\alpha}(\mathbf{k}_{\alpha})\bigg)_{\uparrow\uparrow} &= 2 \,t_{\uparrow\uparrow}^{(\alpha)} \biggr[ \cos(k_{x_{\alpha}}a)+ 2\cos \bigg( \frac{1}{2}k_{x_{\alpha}}a\bigg) \cos\bigg(\frac{\sqrt{3}}{2}k_{y_{\alpha}}a\bigg)  \biggr]-g_{\alpha} M_{\alpha,z},\\
    \bigg(\mathcal{H}_{\alpha}(\mathbf{k}_{\alpha})\bigg)_{\uparrow\downarrow} &= 2i\,t_{\uparrow\downarrow}^{(\alpha)} \biggr[  \sin(k_{x_{\alpha}}a)+ \sin \bigg( \frac{1}{2}k_{x_{\alpha}}a\bigg) \cos\bigg(\frac{\sqrt{3}}{2}k_{y_{\alpha}}a\bigg)-i\sqrt{3}\cos \bigg( \frac{1}{2}k_{x_{\alpha}}a\bigg) \sin\bigg(\frac{\sqrt{3}}{2}k_{y_{\alpha}}a\bigg) \biggr] \\&\hspace{0.2cm}- g_{\alpha}\big(M_{\alpha,x}-i M_{\alpha,y}\big),\\
    \bigg(\mathcal{H}_{\alpha}(\mathbf{k}_{\alpha})\bigg)_{\downarrow\uparrow} &= -2i\,\big(t^{(\alpha)}_{\uparrow \downarrow}\big)^* \biggr[  \sin(k_{x_{\alpha}}a)+ \sin \bigg( \frac{1}{2}k_{x_{\alpha}}a\bigg) \cos\bigg(\frac{\sqrt{3}}{2}k_{y_{\alpha}}a\bigg)+i\sqrt{3}\cos \bigg( \frac{1}{2}k_{x_{\alpha}}a\bigg) \sin\bigg(\frac{\sqrt{3}}{2}k_{y_{\alpha}}a\bigg) \biggr]
     \\&\hspace{0.2cm} - g_{\alpha}\big(M_{\alpha,x}+i M_{\alpha,y}\big),\\
     \bigg(\mathcal{H}_{\alpha}(\mathbf{k}_{\alpha})\bigg)_{\downarrow\downarrow} &= 2 \,t_{\downarrow\downarrow}^{(\alpha)} \biggr[ \cos(k_{x_{\alpha}}a)+ 2\cos \bigg( \frac{1}{2}k_{x_{\alpha}}a\bigg) \cos\bigg(\frac{\sqrt{3}}{2}k_{y_{\alpha}}a\bigg)  \biggr]+g_{\alpha} M_{\alpha,z}.\\
     \end{cases}
\end{align}

\subsection{Analytical calculation of $E^{(2)}_0$ for the square lattice at $\theta = 0$}\label{Section Analytical Calculation}
Although we typically cannot obtain a closed-form expression for $E^{(2)}_0$, it is worthwhile to consider those special cases in which the right-hand side of Eq. \eqref{Equation Second-Order Correction Integral Form} can be evaluated analytically. One such case will be covered in this section.

\subsubsection{Configuration with $\mathbf{M}_1 = \mathbf{M}_2$}
We consider two completely equivalent square lattice layers with $t_{\uparrow\downarrow} = 0$,  $\mathbf{M}_1 = \mathbf{M}_2 = \mathbf{M}$ and $g_1=g_2=g>0$. The twist angle $\theta$ is taken to be zero, and we assume for simplicity that the in-plane component of the displacement vector $\boldsymbol{\Delta}$ between the layers vanishes ($\Delta_x = \Delta_y = 0$). From these assumptions, we immediately have
\begin{align}
    \epsilon_{1,s}(\mathbf{k}) &= \epsilon_{2,s}(\mathbf{k}) = \epsilon_s(\mathbf{k}),\\
    U_1(\mathbf{k}) &= U_2(\mathbf{k}) = U \label{Equation Bravais U-Matrix},
\end{align}
where we note that the $2\times 2$ matrix $U$ depends only on $\mathbf{M}$ here. Furthermore, using Eq. \eqref{Equation Bravais Interlayer Hopping} in conjunction with Eq. \eqref{Equation Bravais U-Matrix}, we find that
\begin{align}
    \Upsilon_{s,s'}(\mathbf{k}_1,\mathbf{k}_2,\mathbf{G}_1,\mathbf{G}_2) &= \frac{1}{\sqrt{\mathcal{A}_1\mathcal{A}_2}} \mathcal{T}^{\mathrm{int}}(\mathbf{G}_1-\mathbf{k}_1) \delta_{ss'},
\end{align}
where $\mathcal{T}^{\mathrm{int}}(\mathbf{G}_1-\mathbf{k}_1)$ is the Fourier transform of the interlayer hopping coefficient $t^{\mathrm{int}}(\mathbf{r})$. Plugging these results into the integral expression for $E^{(2)}_0$, Eq. \eqref{Equation Second-Order Correction Integral Form}, we get
\begin{align}
     E_0^{(2)} &=  \sum_{\mathbf{G}_1,\mathbf{G}_2} \sum_{\mathbf{G_1',\mathbf{G}_2'}} \frac{A}{\mathcal{A}^2} \,\delta_{\mathbf{G}_{12},\mathbf{G}_{12}'} \int_{\Omega_1} \frac{d^2 \mathbf{k}_1}{(2\pi)^2} \Phi_2(\mathbf{k}_1-\mathbf{G}_{12}) \mathcal{T}^{\mathrm{int}}(\mathbf{G}_1-\mathbf{k}_1) \biggr(\mathcal{T}^{\mathrm{int}}(\mathbf{G}_1'-\mathbf{k}_1)\biggr)^* \nonumber\\ &\times  \biggr[\frac{F(\epsilon_{\uparrow}(\mathbf{k}_1-\mathbf{G}_{12}))-F(\epsilon_{\uparrow}(\mathbf{k}_1))}{\epsilon_{\uparrow}(\mathbf{k}_1-\mathbf{G}_{12})-\epsilon_{\uparrow}(\mathbf{k}_1)}+\frac{F(\epsilon_{\downarrow}(\mathbf{k}_1-\mathbf{G}_{12}))-F(\epsilon_{\downarrow}(\mathbf{k}_1))}{\epsilon_{\downarrow}(\mathbf{k}_1-\mathbf{G}_{12})-\epsilon_{\downarrow}(\mathbf{k}_1)}\biggr],
\end{align}
where we have made use of the fact that $\mathcal{A}_1 = \mathcal{A}_2 = \mathcal{A}$ here. Now, because the layers are equivalent and $\theta = 0$, we find that $\Phi_2(\mathbf{k}_1-\mathbf{G}_{12}) = 1$ if $\mathbf{G}_{12} = \mathbf{0}$, and zero otherwise. Using the identity, 
\begin{align}
    \lim_{\Delta \mathbf{k} \rightarrow 0}  \frac{F(\epsilon_{s}(\mathbf{k}+\Delta \mathbf{k}))-F(\epsilon_{s}(\mathbf{k}))}{\epsilon_{s}(\mathbf{k}+\Delta \mathbf{k})-\epsilon_{s}(\mathbf{k})} &= - \delta(\epsilon_s(\mathbf{k})-\epsilon_f),
\end{align}
we then obtain the following expression for $E^{(2)}_0$,
\begin{align}
     E_0^{(2)} &=  -\frac{A}{\mathcal{A}^2}\sum_{\mathbf{G}_1,\mathbf{G}_1'}   \int_{\Omega_1} \frac{d^2 \mathbf{k}_1}{(2\pi)^2} \mathcal{T}^{\mathrm{int}}(\mathbf{G}_1-\mathbf{k}_1) \biggr(\mathcal{T}^{\mathrm{int}}(\mathbf{G}_1'-\mathbf{k}_1)\biggr)^*  \biggr[\delta(\epsilon_{\uparrow}(\mathbf{k}_1)-\epsilon_f)+\delta(\epsilon_{\downarrow}(\mathbf{k}_1)-\epsilon_f)\biggr]. 
\end{align}
To complete our calculations, we need an expression for $\mathcal{T}^{\mathrm{int}}(\mathbf{k})$. While we could in principle obtain this by Fourier transforming Eq. \eqref{Equation Bravais Interlayer Hopping}, we will for the sake of mathematical simplicity work with a simpler expression, namely
\begin{align}
    \mathcal{T}^{\mathrm{int}}(\mathbf{k}) &=
    \begin{cases}
        2\pi\mathcal{A}\, V^{\mathrm{int}}, \hspace{0.1cm} \text{if} \hspace{0.1cm} \mathbf{k} \in \Omega_1.\\
        0, \hspace{0.1cm} \text{otherwise.}
    \end{cases}
\end{align}
We thus effectively treat the interlayer hopping in momentum space as constant over the first Brillouin zone, with a sharp cutoff at the Brillouin zone boundaries. With this approximation, we then have 
\begin{align}\label{Equation Bravais Energy and DOS-like}
    E_0^{(2)} &=  -A|V^{\mathrm{int}}|^2   \int_{\Omega_1} d^2 \mathbf{k}  \biggr[\delta(\epsilon_{\uparrow}(\mathbf{k})-\epsilon_f)+\delta(\epsilon_{\downarrow}(\mathbf{k})-\epsilon_f)\biggr]. 
\end{align}

To evaluate this integral, we need to perform a coordinate transformation in which we express $\mathbf{k}$ as a function of the single-particle energy. More formally, taking $t_{\uparrow \uparrow} = t_{\downarrow \downarrow} = t/2$, the single-particle energies for our square lattice layers are given by
\begin{align}
    \epsilon_{\uparrow}(\mathbf{k}) &= t \biggr[ \cos(k_x a) + \cos(k_y a)\biggr] - g |\mathbf{M}| \equiv \epsilon_t(\mathbf{k})- g |\mathbf{M}| ,\\
    \epsilon_{\downarrow}(\mathbf{k}) &= t \biggr[ \cos(k_x a) + \cos(k_y a)\biggr] + g |\mathbf{M}| \equiv \epsilon_t(\mathbf{k})+ g |\mathbf{M}|.
\end{align} 
Now, in order to perform the coordinate transformation, we first note that $\Omega_1$ consists of all pairs $(k_x,k_y)$ satisfying $-\pi/a < k_x \leq \pi/a$, $-\pi/a < k_y \leq \pi/a$. Due to the symmetry of the integrand, we can focus our attention on the subdomain described by  $0 \leq k_x \leq \pi/a$, $0 \leq k_y \leq \pi/a$, writing
\begin{align}
     E_0^{(2)} &=  -4 A|V^{\mathrm{int}}|^2   \int_{0}^{\pi/a} d k_x \int_{0}^{\pi/a} d k_y  \biggr[\delta(\epsilon_{\uparrow}(\mathbf{k})-\epsilon_f)+\delta(\epsilon_{\downarrow}(\mathbf{k})-\epsilon_f)\biggr].
\end{align}
An invertible coordinate transformation on this subdomain is now given by
\begin{align}
    k_x a &= \arccos\bigg( \frac{\epsilon_t(\mathbf{k})+\xi(\mathbf{k})}{2t}\bigg),\\
    k_y a &= \arccos\bigg( \frac{\epsilon_t(\mathbf{k})-\xi(\mathbf{k})}{2t}\bigg),
\end{align}
where
\begin{align}
    \xi(\mathbf{k}) &= t \biggr[ \cos(k_x a) - \cos(k_y a)\biggr].
\end{align}

\begin{figure}
    \centering
    \includegraphics[scale=0.29]{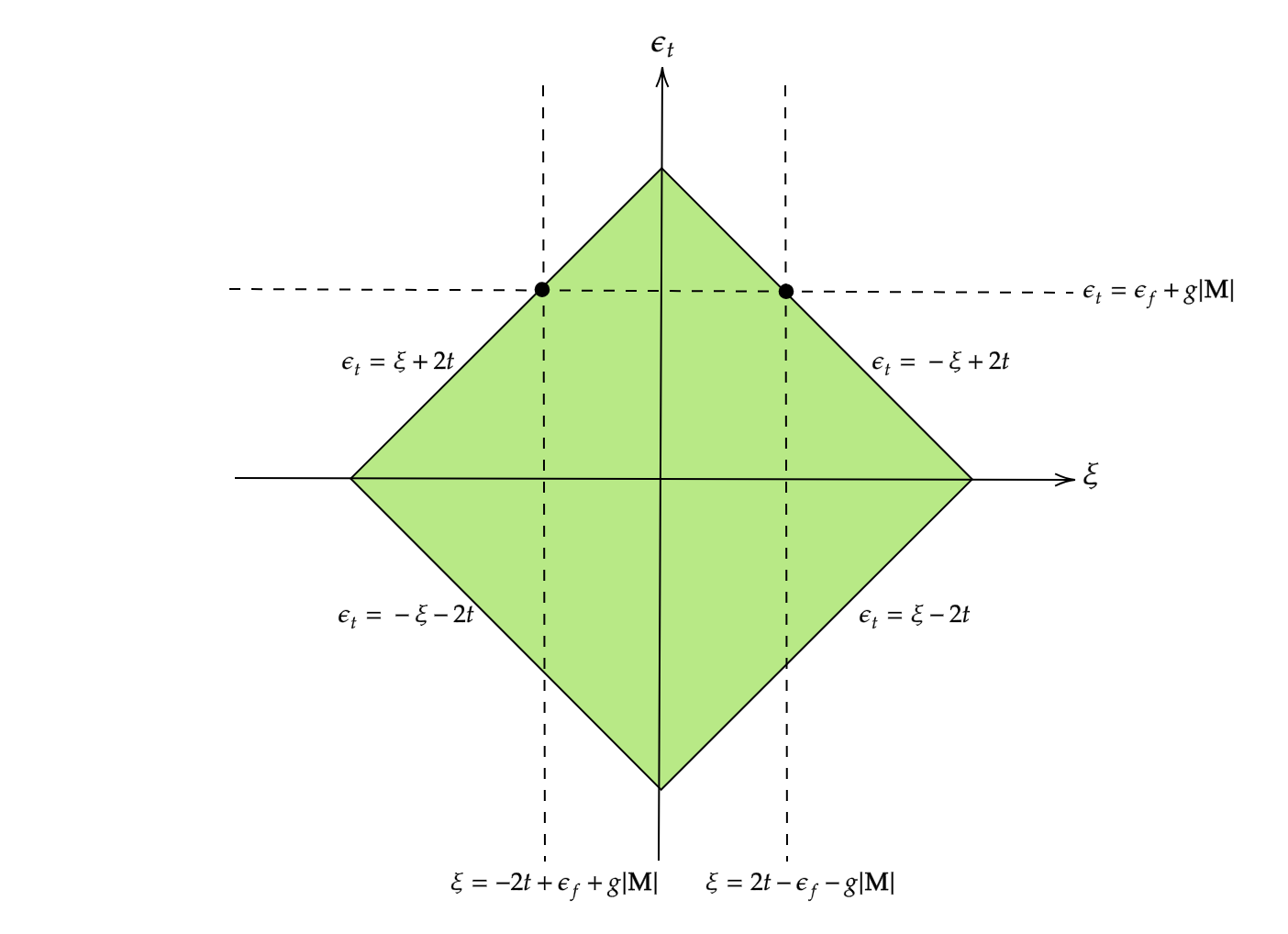}
    \caption{The diamond-shaped integration domain $\mathcal{D}$ in $(\xi,\epsilon_t)$-space. Since our integrands are of the form $ f(\xi,\epsilon_t) \delta(\epsilon_t \pm g|\mathbf{M}|-\epsilon_f)$, and we assume that $-2t <\epsilon_f \pm g|\mathbf{M}|<2t$, the contributions to the integral come solely from the lines $\epsilon_t = \epsilon_f \pm g|\mathbf{M}|$. In this figure we have taken $\epsilon_f + g|\mathbf{M}| \geq 0$, in which case the $\xi$-values of the line $\epsilon_t = \epsilon_f + g|\mathbf{M}|$ that are part of the integration domain run from $\xi=-2t+\epsilon_f+g|\mathbf{M}|$ to  $\xi=2t-\epsilon_f-g|\mathbf{M}|$.}
    \label{Figure Integration}
\end{figure}

The area element $dk_x dk_y$ is related to $d \epsilon_t d \xi$ via
\begin{align}
    d k_x dk_y &= \biggr| \frac{\partial(k_x,k_y)}{\partial(\epsilon_t,\xi)} \biggr| d \epsilon_t d\xi \nonumber \\ &=  \frac{1}{2 a^2t^2} \biggr\{ \biggr[1-\biggr(\frac{\epsilon_t+\xi}{2t}\biggr)^2\biggr]\biggr[1-\biggr(\frac{\epsilon_t-\xi}{2t}\biggr)^2\biggr] \biggr\}^{-1/2} d \epsilon_t d\xi,
\end{align}
while the transformed integration domain can be found by noting that $\cos(k_x a)$ and $\cos(k_y a)$ are invertible functions on our $\mathbf{k}$-space subdomain, which take on all values in the range $[-1,1]$. Since $\cos(k_x a) = (\epsilon_t+\xi)/2t$ and $\cos(k_y a) = (\epsilon_t-\xi)/2t$, the integration domain $\mathcal{D}$ in the $(\xi,\epsilon_t)$-plane is a diamond-shaped square consisting of all pairs $(\xi,\epsilon_t)$ that satisfy the inequalities $-2 t \leq \epsilon_t+\xi \leq 2 t$ and $-2 t \leq \epsilon_t-\xi \leq 2 t$. The domain $\mathcal{D}$ is drawn in Fig. \ref{Figure Integration}. 

In terms of the new coordinates, the integral now takes on the following form,
\begin{align}
    E_0^{(2)} &=  -\frac{2A |V^{\mathrm{int}}|^2}{a^2t^2} \int_{\mathcal{D}} d\xi \, d \epsilon_t \, \biggr\{ \biggr[1-\biggr(\frac{\epsilon_t+\xi}{2t}\biggr)^2\biggr]\biggr[1-\biggr(\frac{\epsilon_t-\xi}{2t}\biggr)^2\biggr] \biggr\}^{-1/2} \biggr[\delta(\epsilon_t-g|\mathbf{M}|-\epsilon_f)+\delta(\epsilon_t+g|\mathbf{M}|-\epsilon_f)\biggr]. 
\end{align}
Since under experimental conditions one almost always has both $g|\mathbf{M}| < t$ and $|\epsilon_f| < t$, we focus our attention on the case where $-2t < \epsilon_f - g|\mathbf{M}| < \epsilon_f + g|\mathbf{M}| < 2 t$. Due to the Dirac delta functions in the integrand, the integral over $\epsilon_t$ can then readily be performed. We must, however, be careful with where the energies $\epsilon_f \pm g|\mathbf{M}|$ are located inside $\mathcal{D}$, and we therefore distinguish between three possible situations: 
\begin{enumerate}
    \item \textbf{Case I}: $\epsilon_f + g|\mathbf{M}| \geq 0$, $\epsilon_f - g|\mathbf{M}| \geq 0$.\\ In this case, we find that the $\xi$-values of the line $\epsilon_t = \epsilon_f + g|\mathbf{M}|$ that are part of the integration domain run from $\xi=-2t+\epsilon_f+g|\mathbf{M}|$ to $\xi=2t-\epsilon_f-g|\mathbf{M}|$. Similarly, the $\xi$-values of the line $\epsilon_t = \epsilon_f - g|\mathbf{M}|$ that are part of the integration domain run from $\xi=-2t+\epsilon_f-g|\mathbf{M}|$ to $\xi=2t-\epsilon_f+g|\mathbf{M}|$. Observing that the integrand is symmetric around $\xi = 0$, we then find
    \begin{align}\label{Equation Bravais Integrals Case I}
         E_0^{(2)} &=  -\frac{4A |V^{\mathrm{int}}|^2}{a^2t^2} \int_0^{2t-\epsilon_f-g|\mathbf{M}|} d \xi \, \biggr\{ \biggr[1-\biggr(\frac{\epsilon_f+g|\mathbf{M}|+\xi}{2t}\biggr)^2\biggr]\biggr[1-\biggr(\frac{\epsilon_f+g|\mathbf{M}|-\xi}{2t}\biggr)^2\biggr] \biggr\}^{-1/2} \nonumber \\&-\frac{4A |V^{\mathrm{int}}|^2}{a^2t^2} \int_0^{2t-\epsilon_f+g|\mathbf{M}|} d \xi \, \biggr\{ \biggr[1-\biggr(\frac{\epsilon_f-g|\mathbf{M}|+\xi}{2t}\biggr)^2\biggr]\biggr[1-\biggr(\frac{\epsilon_f-g|\mathbf{M}|-\xi}{2t}\biggr)^2\biggr] \biggr\}^{-1/2}.
    \end{align}
    \item \textbf{Case II}: $\epsilon_f + g|\mathbf{M}| \geq 0$, $\epsilon_f - g|\mathbf{M}| \leq 0$.\\
    In this case, we find that the $\xi$-values of the line $\epsilon_t = \epsilon_f + g|\mathbf{M}|$ that are part of the integration domain run from $\xi=-2t+\epsilon_f+g|\mathbf{M}|$ to $\xi=2t-\epsilon_f-g|\mathbf{M}|$. Similarly, the $\xi$-values of the line $\epsilon_t = \epsilon_f - g|\mathbf{M}|$ that are part of the integration domain run from $\xi=-2t-\epsilon_f+g|\mathbf{M}|$ to $\xi=2t+\epsilon_f-g|\mathbf{M}|$. Observing that the integrand is symmetric around $\xi = 0$, we then find
    \begin{align}
     E_0^{(2)} &=  -\frac{4A |V^{\mathrm{int}}|^2}{a^2t^2} \int_0^{2t-\epsilon_f-g|\mathbf{M}|} d \xi \, \biggr\{ \biggr[1-\biggr(\frac{\epsilon_f+g|\mathbf{M}|+\xi}{2t}\biggr)^2\biggr]\biggr[1-\biggr(\frac{\epsilon_f+g|\mathbf{M}|-\xi}{2t}\biggr)^2\biggr] \biggr\}^{-1/2} \nonumber \\&-\frac{4A |V^{\mathrm{int}}|^2}{a^2t^2} \int_0^{2t+\epsilon_f-g|\mathbf{M}|} d \xi \, \biggr\{ \biggr[1-\biggr(\frac{\epsilon_f-g|\mathbf{M}|+\xi}{2t}\biggr)^2\biggr]\biggr[1-\biggr(\frac{\epsilon_f-g|\mathbf{M}|-\xi}{2t}\biggr)^2\biggr] \biggr\}^{-1/2}.
    \end{align}
    \item \textbf{Case III}: $\epsilon_f + g|\mathbf{M}| \leq 0$, $\epsilon_f - g|\mathbf{M}| \leq 0$.\\
    In this case, we find that the $\xi$-values of the line $\epsilon_t = \epsilon_f + g|\mathbf{M}|$ that are part of the integration domain run from  $\xi=-2t-\epsilon_f-g|\mathbf{M}|$ to $\xi=2t+\epsilon_f+g|\mathbf{M}|$. Similarly, the $\xi$-values of the line $\epsilon_t = \epsilon_f - g|\mathbf{M}|$ that are part of the integration domain run from $\xi=-2t-\epsilon_f+g|\mathbf{M}|$ to $\xi=2t+\epsilon_f-g|\mathbf{M}|$. Observing that the integrand is symmetric around $\xi = 0$, we then find
    \begin{align}\label{Equation Bravais Integrals Case III}
         E_0^{(2)} &=  -\frac{4A |V^{\mathrm{int}}|^2}{a^2t^2} \int_0^{2t+\epsilon_f+g|\mathbf{M}|} d \xi \, \biggr\{ \biggr[1-\biggr(\frac{\epsilon_f+g|\mathbf{M}|+\xi}{2t}\biggr)^2\biggr]\biggr[1-\biggr(\frac{\epsilon_f+g|\mathbf{M}|-\xi}{2t}\biggr)^2\biggr] \biggr\}^{-1/2} \nonumber \\&-\frac{4A |V^{\mathrm{int}}|^2}{a^2t^2} \int_0^{2t+\epsilon_f-g|\mathbf{M}|} d \xi \, \biggr\{ \biggr[1-\biggr(\frac{\epsilon_f-g|\mathbf{M}|+\xi}{2t}\biggr)^2\biggr]\biggr[1-\biggr(\frac{\epsilon_f-g|\mathbf{M}|-\xi}{2t}\biggr)^2\biggr] \biggr\}^{-1/2}.
    \end{align}
\end{enumerate}

Only one of the four different integrals in the above expressions is unique. This follows from the fact that the integrals of Eq.~\eqref{Equation Bravais Integrals Case I} transform into those of Eq.~\eqref{Equation Bravais Integrals Case III} when we take $\epsilon_f \pm g|\mathbf{M}| \rightarrow -(\epsilon_f \pm g|\mathbf{M}| )$, while the integrals of Eq.~\eqref{Equation Bravais Integrals Case I} itself transform into one another when we take $g|\mathbf{M}| \rightarrow -g|\mathbf{M}|$. We thus only tackle one of these integrals, which we will refer to with $I$. Defining $\bar{\xi} = \frac{\xi}{2t}$ and $\bar{\epsilon} = \frac{\epsilon_f+g|\mathbf{M}|}{2t}$, we get
\begin{align}
         I(\epsilon_f+g|\mathbf{M}|) &=  -\frac{4A |V^{\mathrm{int}}|^2}{a^2t^2} \int_0^{2t-\epsilon_f-g|\mathbf{M}|} d \xi \, \biggr\{ \biggr[1-\biggr(\frac{\epsilon_f+g|\mathbf{M}|+\xi}{2t}\biggr)^2\biggr]\biggr[1-\biggr(\frac{\epsilon_f+g|\mathbf{M}|-\xi}{2t}\biggr)^2\biggr] \biggr\}^{-1/2} \nonumber \\
         &= -\frac{8A |V^{\mathrm{int}}|^2}{a^2t} \int_0^{1-\bar{\epsilon}} d \bar{\xi} \, \biggr\{ \biggr[1-\big(\bar{\epsilon}+\bar{\xi}\big)^2\biggr]\biggr[1-\big(\bar{\epsilon}-\bar{\xi}\big)^2\biggr] \biggr\}^{-1/2} \nonumber \\
         &= -\frac{8A |V^{\mathrm{int}}|^2}{a^2t} \int_0^{1-\bar{\epsilon}} d \bar{\xi} \, \biggr\{ \big(1+\bar{\epsilon}-\bar{\xi}\big)\big(1-\bar{\epsilon}-\bar{\xi}\big)\big[\bar{\xi}-(-1+\bar{\epsilon})\big]\big[\bar{\xi}-(-1-\bar{\epsilon})\big]\biggr\}^{-1/2}.
    \end{align}
When $0<\bar{\epsilon} < 1$ (or equivalently, $0<\epsilon_f+g|\mathbf{M}|<2t$), we can make use of the following identity \cite{Byrd1971}, 
\begin{align}
    \int_y^b \frac{d x}{\sqrt{(a-x)(b-x)(x-c)(x-d)}} &= \frac{2}{\sqrt{(a-c)(b-d)}} \mathcal{F}\biggr( \arcsin\biggr(\sqrt{\frac{(a-c)(b-y)}{(b-c)(a-y)}}\biggr), \frac{(b-c)(a-d)}{(a-c)(b-d)} \biggr),
\end{align}
with $a>b>y\geq c >d$, and with $\mathcal{F}(\phi,k^2)$ the incomplete elliptic integral of the first kind, given by
\begin{align}
    \mathcal{F}(\phi,k^2) &= \int_0^{\phi} \frac{d \psi}{\sqrt{1-k^2\sin^2 (\psi)}}.
\end{align}
We then obtain
\begin{align}
   I(\epsilon_f+g|\mathbf{M}|) &=  -\frac{8 A |V^{\mathrm{int}}|^2}{a^2t} \mathcal{F}\biggr( \arcsin\biggr(\frac{1}{\sqrt{1+\frac{\epsilon_f+g|\mathbf{M}|}{2t}}}\biggr), \bigg(1-\frac{(\epsilon_f+g|\mathbf{M}|)}{2t} \bigg)\bigg(1+\frac{\epsilon_f+g|\mathbf{M}|}{2t} \bigg) \biggr).
\end{align}
When $\bar{\epsilon} = 0$ (or equivalently, $\epsilon_f+g|\mathbf{M}|=0$), the integral $I$ diverges. This can be understood immediately by realizing that the right-hand side of Eq. \eqref{Equation Bravais Energy and DOS-like} is proportional to the density of states (DOS) of the unperturbed system. The divergence of $I$ is thus a simple consequence of the fact that the van Hove singularities at $\epsilon=0$ lead to a logarithmic divergence in the DOS \cite{VanHove1953}.

Having finished the calculation of $I$, we can thus conclude this section by stating the resulting energy corrections for the three cases introduced previously. All of these cases can be summarized in one single equation, which is given by
\begin{align}
    E^{(2)}_0 &= I(\big|\,\epsilon_f+g|\mathbf{M}|\,\big|)+I(\big|\,\epsilon_f-g|\mathbf{M}|\,\big|). 
\end{align}
Here, we have assumed that neither $\epsilon_f + g|\mathbf{M}|$ nor $\epsilon_f - g|\mathbf{M}|$ is zero. When either or both equal zero, the energy correction $E^{(2)}_0$ diverges.


\subsubsection{Configuration with $\mathbf{M}_1 = \mathbf{M}$ and $\mathbf{M}_2 = \mathbf{0}$}
We consider a similar configuration as in the previous section, the only difference being that we now take $\mathbf{M}_1 = \mathbf{M}$ and $\mathbf{M}_2 = \mathbf{0}$. Since we can now freely choose an orthogonal spin space eigenbasis for the second layer, we can choose $U_2$ such that we again have
\begin{align}
    U_1(\mathbf{k}) = U_2(\mathbf{k}) = U.
\end{align}
Using similar arguments as in the previous section, we now find
\begin{align}
     E_0^{(2)} &= A|V^{\mathrm{int}}|^2  \int_{\Omega_1} d^2 \mathbf{k}\,   \biggr[\frac{F(\epsilon_t(\mathbf{k}))-F(\epsilon_t(\mathbf{k})-g|\mathbf{M}|)}{g|\mathbf{M}|}+\frac{F(\epsilon_t(\mathbf{k}))-F(\epsilon_t(\mathbf{k})+g|\mathbf{M}|)}{-g|\mathbf{M}|}\biggr] \nonumber \\&= \frac{A|V^{\mathrm{int}}|^2}{ g |\mathbf{M}|} \int_{\Omega_1} d^2 \mathbf{k}\, \biggr[ F(\epsilon_t(\mathbf{k})+g|\mathbf{M}|) - F(\epsilon_t(\mathbf{k})-g|\mathbf{M}|)     \biggr].
\end{align}
Using the same coordinate transformation as before, we obtain the following expression for $E_0^{(2)}$,
\begin{align}
     E_0^{(2)} &= \frac{2A |V^{\mathrm{int}}|^2}{a^2t^2g|\mathbf{M}|} \int_{\mathcal{D}} d \xi \, d \epsilon_t \, \biggr\{ \biggr[1-\biggr(\frac{\epsilon_t+\xi}{2t}\biggr)^2\biggr]\biggr[1-\biggr(\frac{\epsilon_t-\xi}{2t}\biggr)^2\biggr] \biggr\}^{-1/2}\biggr[ F(\epsilon_t+g|\mathbf{M}|) - F(\epsilon_t-g|\mathbf{M}|)     \biggr],
\end{align}
where we observe that
\begin{align}
     F(\epsilon_t+g|\mathbf{M}|) - F(\epsilon_t-g|\mathbf{M}|) &=
     \begin{cases}
         -1,\hspace{0.1cm} \text{for} \hspace{0.1cm} \epsilon_f - g|\mathbf{M}| \leq \epsilon_t < \epsilon_f + g|\mathbf{M}|.\\
         0, \hspace{0.1cm} \text{otherwise}.
     \end{cases}
\end{align}
Now, assuming $-2t < \epsilon_f - g|\mathbf{M}| < \epsilon_f + g|\mathbf{M}| < 2 t$, we immediately obtain
\begin{align}
    E_0^{(2)} &= -\frac{4A |V^{\mathrm{int}}|^2}{a^2t^2g|\mathbf{M}|} \int_{\epsilon_f-g|\mathbf{M}|}^{\epsilon_f+g|\mathbf{M}|} d \epsilon_t \int_0^{\chi(\epsilon_t)} d \xi \, \biggr\{ \biggr[1-\biggr(\frac{\epsilon_t+\xi}{2t}\biggr)^2\biggr]\biggr[1-\biggr(\frac{\epsilon_t-\xi}{2t}\biggr)^2\biggr] \biggr\}^{-1/2},
\end{align}
where $\chi(\epsilon)$ is the $\xi$-coordinate of the point of intersection between $\epsilon_t = \epsilon$ and the right boundary of $\mathcal{D}$. Introducing the following definitions,
\begin{align}
    \epsilon_{\mathrm{max}} &= \max(\big|\,\epsilon_f+g|\mathbf{M}|\,\big|,\big|\,\epsilon_f-g|\mathbf{M}|\,\big|),\\
    \epsilon_{\mathrm{min}} &= \min(\big|\,\epsilon_f+g|\mathbf{M}|\,\big|,\big|\,\epsilon_f-g|\mathbf{M}|\,\big|),
\end{align}
we find that the energy correction is given by
\begin{align}
    E^{(2)}_0 &= \frac{1}{g|\mathbf{M}|}\int_{\epsilon_{\mathrm{min}}}^{\epsilon_{\mathrm{max}}} d \epsilon_t \, I(\epsilon_t),
\end{align}
when $\epsilon_f+g|\mathbf{M}|$ and $\epsilon_f-g|\mathbf{M}|$ have the same sign. When $\epsilon_f+g|\mathbf{M}| \geq 0$ and $\epsilon_f-g|\mathbf{M}| \leq 0$ we find instead
\begin{align}
    E^{(2)}_0 &= \frac{1}{g|\mathbf{M}|}\biggr[\int_{0}^{\epsilon_{\mathrm{min}}} d \epsilon_t \, I(\epsilon_t) + \int_{0}^{\epsilon_{\mathrm{max}}} d \epsilon_t \, I(\epsilon_t)\biggr].
\end{align}
Finally, we note that one arrives at exactly the same results if one would have started with $\mathbf{M}_1 = \mathbf{0}$ and $\mathbf{M}_2 = \mathbf{M}$.

\subsubsection{Configuration with $\mathbf{M}_1 \perp \mathbf{M}_2$}
We consider a similar configuration as in the previous section, the only difference being that we now take $\mathbf{M}_1 \perp \mathbf{M}_2$ and $|\mathbf{M}_1| = |\mathbf{M}_2| = |\mathbf{M}|$. We will focus on the case where $\mathbf{M}_1$ and $\mathbf{M}_2$ point along the Cartesian coordinate axes, as all other possible situations can be derived from this one by means of suitable rotations.

For a given $\mathbf{M}$, the matrix $U$ is chosen such that $U^{\dagger} \big( \mathbf{M} \cdot \mathbf{S} \big) U = |\mathbf{M}| \sigma_z$. We then obtain the following relations,
\begin{align}
    \begin{cases}
    \mathbf{M} &= \pm |\mathbf{M}| \hat{\mathbf{x}} \rightarrow U = e^{\mp i\frac{\pi}{4} \sigma_y},\\
    \mathbf{M} &= \pm |\mathbf{M}| \hat{\mathbf{y}} \rightarrow U = e^{\pm i\frac{\pi}{4} \sigma_x},\\
    \mathbf{M} &= \pm |\mathbf{M}| \hat{\mathbf{z}} \rightarrow U = e^{\mp i\frac{\pi}{4} \sigma_z},
    \end{cases}
\end{align}
with $\{\sigma_x,\sigma_y,\sigma_z\}$ the Pauli matrices. Because now $U_1^{\dagger}U_2 \neq I$, we have to modify the expression for $\Upsilon^{\mathrm{int}}_{s,s'}$ that we used in the previous sections to
\begin{align}
    \Upsilon^{\mathrm{int}}_{s,s'}(\mathbf{k}_1,\mathbf{k}_2,\mathbf{G}_1,\mathbf{G}_2) &= \frac{1}{\sqrt{\mathcal{A}_1\mathcal{A}_2}} \mathcal{T}^{\mathrm{int}}(\mathbf{G}_1-\mathbf{k}_1)\biggr(U_1^{\dagger} U_2\biggr)_{s,s'}.
\end{align}
Plugging this into the integral formula for $E^{(2)}_0$ and making use of the fact that $\epsilon_{1,s}(\mathbf{k}) = \epsilon_{2,s}(\mathbf{k})$ here, we immediately find that the energy correction for this type of configuration is given by
\begin{align}
     E_0^{(2)} &=  A|V^{\mathrm{int}}|^2  \int_{\Omega_1} d^2 \mathbf{k}   \sum_{s,s}\biggr|\biggr( U_1^{\dagger} U_2\biggr)_{ss'}\biggr|^2\biggr[\frac{F(\epsilon_{s'}(\mathbf{k}))-F(\epsilon_{s}(\mathbf{k}))}{\epsilon_{s'}(\mathbf{k})-\epsilon_{s}(\mathbf{k})}\biggr] \nonumber\\
     &= -A|V^{\mathrm{int}}|^2 \biggr|\biggr( U_1^{\dagger} U_2\biggr)_{\uparrow \uparrow}\biggr|^2  \int_{\Omega_1} d^2 \mathbf{k} \, \delta(\epsilon_{\uparrow}(\mathbf{k})-\epsilon_f)-A|V^{\mathrm{int}}|^2 \biggr|\biggr( U_1^{\dagger} U_2\biggr)_{\downarrow \downarrow}\biggr|^2 \int_{\Omega_1} d^2 \mathbf{k}\, \nonumber \delta(\epsilon_{\downarrow}(\mathbf{k})-\epsilon_f) \nonumber \\&+A|V^{\mathrm{int}}|^2 \biggr(\,\biggr|\biggr( U_1^{\dagger} U_2\biggr)_{\uparrow \downarrow}\biggr|^2+\biggr|\biggr( U_1^{\dagger} U_2\biggr)_{\downarrow \uparrow}\biggr|^2\biggr)  \int_{\Omega_1} d^2 \mathbf{k} \biggr[\frac{F(\epsilon_t(\mathbf{k})+g|\mathbf{M}|)-F(\epsilon_t(\mathbf{k})-g|\mathbf{M}|)}{2 g |\mathbf{M}|}\biggr].
\end{align}
Using the results of the previous sections, we can immediately perform the integrals in the above expression, thus finding
\begin{align}\label{Equation Bravais Energy Correction Orthogonal 1}
    E_0^{(2)} &= \biggr|\biggr( U_1^{\dagger} U_2\biggr)_{\uparrow \uparrow}\biggr|^2 I(\big|\,\epsilon_f+g|\mathbf{M}|\,\big|)+\biggr|\biggr( U_1^{\dagger} U_2\biggr)_{\downarrow \downarrow}\biggr|^2 I(\big|\,\epsilon_f-g|\mathbf{M}|\,\big|) \nonumber \\&+\frac{1}{2 g|\mathbf{M}|} \biggr(\, \biggr|\biggr(U_1^{\dagger} U_2\biggr)_{\uparrow \downarrow}\biggr|^2 +\biggr|\biggr( U_1^{\dagger} U_2\biggr)_{\downarrow \uparrow}\biggr|^2\biggr)\int_{\epsilon_{\mathrm{min}}}^{\epsilon_{\mathrm{max}}} d \epsilon_t \, I(\epsilon_t),
\end{align}
when $\epsilon_f+g|\mathbf{M}|$ and $\epsilon_f-g|\mathbf{M}|$ have the same sign. When $\epsilon_f+g|\mathbf{M}| \geq 0$ and $\epsilon_f-g|\mathbf{M}| \leq 0$ we find instead
\begin{align}\label{Equation Bravais Energy Correction Orthogonal 2}
    E_0^{(2)} &= \biggr|\biggr( U_1^{\dagger} U_2\biggr)_{\uparrow \uparrow}\biggr|^2 I(\big|\,\epsilon_f+g|\mathbf{M}|\,\big|)+\biggr|\biggr( U_1^{\dagger} U_2\biggr)_{\downarrow \downarrow}\biggr|^2I(\big|\,\epsilon_f-g|\mathbf{M}|\,\big|) \nonumber \\&+\frac{1}{2 g|\mathbf{M}|} \biggr(\, \biggr|\biggr(U_1^{\dagger} U_2\biggr)_{\uparrow \downarrow}\biggr|^2 +\biggr|\biggr( U_1^{\dagger} U_2\biggr)_{\downarrow \uparrow}\biggr|^2\biggr)\biggr[\int_{0}^{\epsilon_{\mathrm{min}}} d \epsilon_t \, I(\epsilon_t) + \int_{0}^{\epsilon_{\mathrm{max}}} d \epsilon_t \, I(\epsilon_t)\biggr].
\end{align}
Note that for both of these expressions, we have assumed that neither $\epsilon_f + g|\mathbf{M}|$ nor $\epsilon_f - g|\mathbf{M}|$ is zero. When either or both equal zero, the energy correction diverges. Finally, we mention that the validity of Eqs.~\eqref{Equation Bravais Energy Correction Orthogonal 1} and \eqref{Equation Bravais Energy Correction Orthogonal 2} is not restricted to orthogonal configurations, as the derivation only depended on the equality $|\mathbf{M}_1| = |\mathbf{M}_2| = |\mathbf{M}|$.

\section{Numerical Results for the Bilayer Square Lattice System}\label{Section Numerical Results for the Bilayer Square Lattice System}

In this section, we collect all the results we have obtained by numerically calculating $E^{(2)}_0$ from Eq.~\eqref{Equation Second-Order Correction Integral Form} and extracting the magnetic couplings $J$, $\mathbf{D}$ and $\mathcal{P}$. We begin by demonstrating the excellent agreement with the analytical results for $\theta=0$, as derived in the previous section, and then compare the results obtained with the matrix method with those obtained using the method of least squares. Finally, we present plots of $J$, $\mathbf{D}$ and $\mathcal{P}$ for a variety of twist angles.

\subsection{Comparison with analytical results for $\theta = 0$}
\begin{figure}
    \centering
    \includegraphics[scale=1]{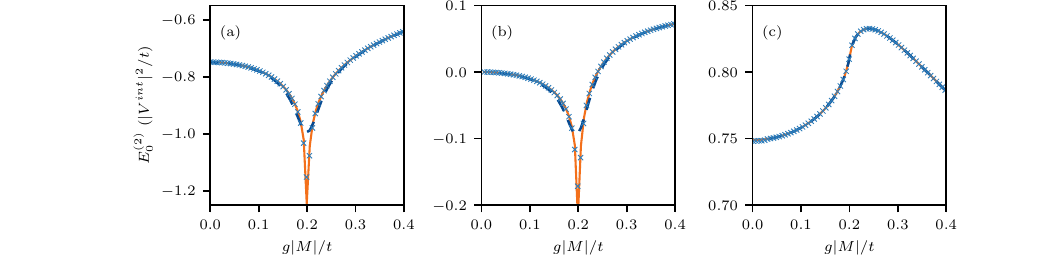}
    \caption{Plots of $E^{(2)}_0$ at $\theta=0$ and $\epsilon_f = 0.2 t$ for different exchange field configurations as a function of the exchange splitting $g|\mathbf{M}|$. The dark blue dashed line and the light blue crosses corresponds to the result obtained by a numerical integration of Eq.~\eqref{Equation Second-Order Correction Integral Form} with finite temperatures $k_B T =  10^{-2}\, t$ and $k_B T =  2\times10^{-3}\, t$, respectively, while the orange line corresponds to the analytical results derived in Section \ref{Section Analytical Calculation}. The following parameters were used: $A/a^2 = 1$, $t= 1$ (arb. units), $V^{\mathrm{int}} = t/2 \pi$. (a) Plot for the configuration $\mathbf{M}_1 = \mathbf{M}_2 = \pm |\mathbf{M}| \hat{\mathbf{x}}$. (b) Plot for the configuration $\mathbf{M}_1 = |\mathbf{M}| \hat{\mathbf{x}}$ and $\mathbf{M}_2 = |\mathbf{M}| \hat{\mathbf{y}}$. (c) Plot for the configuration $\mathbf{M}_1 = |\mathbf{M}| \hat{\mathbf{x}}$ and $\mathbf{M}_2 = \mathbf{0}$. }
    \label{Figure Analytical vs Numerical}
\end{figure}

Here, we present plots of the energy correction $E_0^{(2)}$ for the bilayer square lattice system with $\theta=0$, $t_{\uparrow\downarrow} = 0$, $\Delta_x = \Delta_y = 0$, and 
\begin{align}
    \mathcal{T}^{\mathrm{int}}(\mathbf{k}) &=
    \begin{cases}
        2\pi\mathcal{A}\, V^{\mathrm{int}}, \hspace{0.1cm} \text{if} \hspace{0.1cm} \mathbf{k} \in \Omega_1.\\
        0, \hspace{0.1cm} \text{otherwise.}
    \end{cases}
\end{align}
As shown in Fig. \ref{Figure Analytical vs Numerical}, the results obtained by a numerical integration of Eq.~\eqref{Equation Second-Order Correction Integral Form} are in excellent agreement with the analytical results derived in Section \ref{Section Analytical Calculation}. Although the peaks observed in the numerically computed energy corrections are less sharp compared to the corresponding analytical results, we stress here that this artifact arises from the finite-temperature approximation used in the numerical evaluation. We observe that a reduction of the temperature from $k_B T = 10^{-2}t$ to $k_B T = 2\times10^{-3}t$ already yields a significantly better agreement with the analytical results in the region $g|\mathbf{M}| \approx \epsilon_f$. A drawback of smaller temperatures, however, is that they can introduce numerical instabilities in the integration, and we therefore mostly work with the temperature $k_B T = 10^{-2}t$ in this paper.

\subsection{Comparison between the matrix method and the method of least squares}
Here, we present plots of several quantities of relevance, most notable the Heisenberg exchange constant $J$ and the  DMI vector magnitude $|\mathbf{D}|$, as determined via both the matrix method and the method of least squares.

\begin{figure}
    \centering
\includegraphics[scale=1]{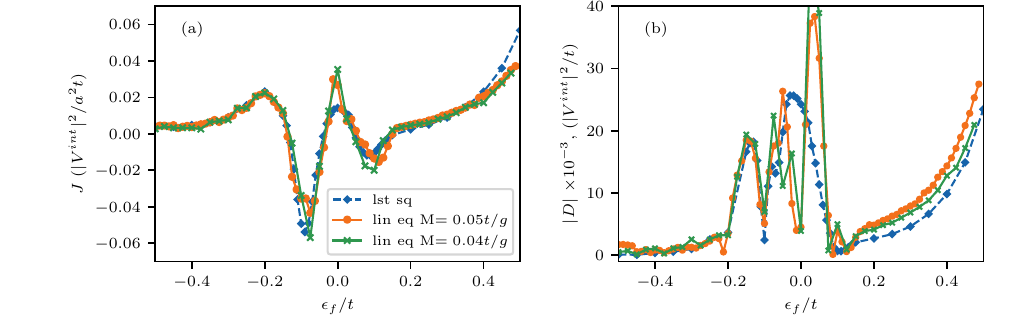}        
    \caption{Plots of (a) J and (b) $|\mathbf{D}|$ as obtained via two different methods --- the matrix method (\textit{lin eq}) and the method of least squares (\textit{lst sq}) --- for twist angle $\theta = \pi/10$ ($18^{\circ}$). For the method of least squares, we used 800 different exchange field configurations, randomly selected and with the magnitudes of $\mathbf{M}_1$ and $\mathbf{M}_2$ ranging from $0.001 \, t/g$ to $0.1 \, t/g$. For the matrix method, magnitudes $|\mathbf{M}|=0.04 \,t/g$ and $|\mathbf{M}|=0.05 \, t/g$ were used. The following parameters were used: t = 1 (arb. units), $t_{\uparrow \downarrow} = 0.1 t$, $a=\Delta_z = r_0 = 1$ (arb. units).}
\label{Figure Matrix vs Least Squares}
\end{figure}

The plots are presented in Fig. \ref{Figure Matrix vs Least Squares}. We find that the results obtained from the matrix method and the method of least squares show excellent agreement for the interlayer Heisenberg exchange coupling at the particular twist angle considered. For the DMI, however, we find slight discrepancies between the two methods for certain values of the chemical potential/Fermi energy. We attribute this mismatch to the relatively small magnitude of the DMI vector in our system, which makes it more sensitive to numerical integration errors. These errors particularly affect the matrix method, which is more susceptible to such inaccuracies than the least-squares approach. We note that higher values for the DMI can be achieved by increasing the spin-flip hopping coefficient $t_{\uparrow\downarrow}$.

For the sake of compactness, we have only presented plots of $J$ and $|\mathbf{D}|$ here, but we emphasize that similar results have been obtained for the other interlayer magnetic coupling coefficients.

\subsection{Plots of $J$, $\mathbf{D}$ and $\mathcal{P}$ for a variety of twist angles $\theta$}
\begin{figure}
    \centering
    \includegraphics[scale=1]{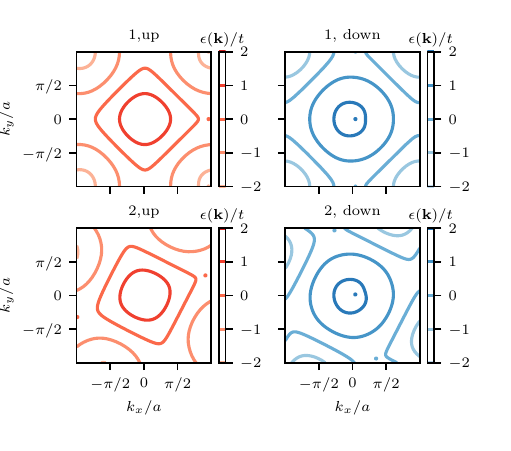}
    \caption{Plots of the energy bands $\epsilon_{1,\uparrow}(\mathbf{k})$, $\epsilon_{1,\downarrow}(\mathbf{k})$, $\epsilon_{2,\uparrow}(\mathbf{k})$ and $\epsilon_{2,\downarrow}(\mathbf{k})$ for the square lattice. The second layer is rotated by an angle $\theta = \pi/10$ ($18^{\circ}$) relative to the first layer. The following parameters were used: $t= 1$ (arb. units), $t_{\uparrow\downarrow} = 0.2t$, and $\mathbf{M}_1 = \mathbf{M}_2 = |\mathbf{M}| \hat{\mathbf{x}}$, with $|\mathbf{M}| = 0.05t/g$. Colors from righter at negative energies to darker at positive energies indicate constant energy levels.}
    \label{Figure Rotated Dispersions}
\end{figure}

We already mentioned in the main body of the paper that the twist-angle dependence of $E^{(2)}_0$ originates from the relative rotation of the two Brillouin zones, which affects both Umklapp scattering ---via rotated reciprocal lattice vectors --- and the single-layer band energies through the rotation of the underlying dispersion. Plots of such rotated dispersion relations for a bilayer square lattice system at relative twist angle $\theta = \pi/10$ are shown in Fig. \ref{Figure Rotated Dispersions}.

\begin{figure}
    \centering
    \includegraphics[scale=1.0]{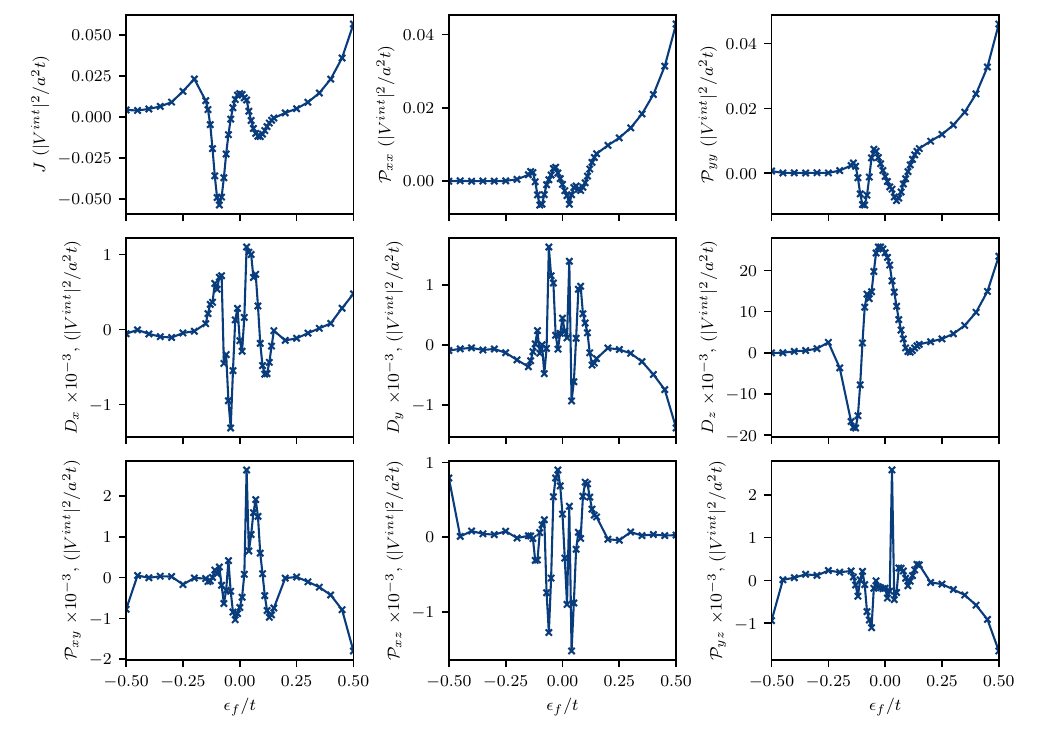}
    \caption{Plots of the interlayer Heisenberg exchange constant $J$, the  DMI vector $\mathbf{D}$, and the five independent components of the traceless, symmetric anistropy matrix $\mathcal{P}$ as functions of the Fermi energy $\epsilon_f$ for the twist angle $\theta = \pi/10$ ($18^{\circ}$). The interlayer hopping is of the Slater-Koster type, $t^{\mathrm{int}}_{\sigma \sigma'}(\mathbf{r}) = V^{\mathrm{int}}  \exp(- r/r_0) \delta_{\sigma \sigma'}$.  The following parameters were used: t = 1 (arb. units), $t_{\uparrow \downarrow} = 0.1t$, $a=\Delta_z = r_0 = 1$ (arb. units).}
    \label{Figure: all-interactions-9panel-pi-10}
\end{figure}

\begin{figure}
    \centering
    \includegraphics[scale=1.0]{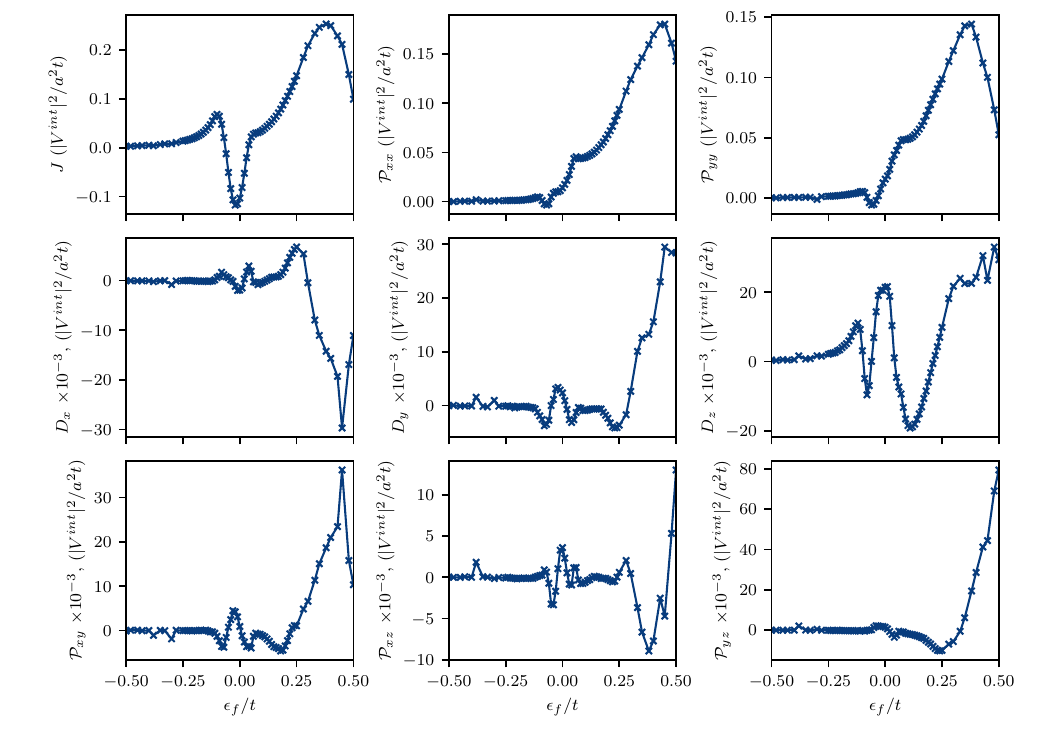}
    \caption{Plots of the interlayer Heisenberg exchange constant $J$, the DMI vector $\mathbf{D}$, and the five independent components of the traceless, symmetric anistropy matrix $\mathcal{P}$ as functions of the Fermi energy $\epsilon_f$ for the twist angle $\theta = \pi/20$ ($9^{\circ}$). The interlayer hopping is of the Slater-Koster type, $t^{\mathrm{int}}_{\sigma \sigma'}(\mathbf{r}) = V^{\mathrm{int}}  \exp(- r/r_0) \delta_{\sigma \sigma'}$.  The following parameters were used: t = 1 (arb. units), $t_{\uparrow \downarrow} = 0.1t$, $a=\Delta_z = r_0 = 1$ (arb. units).}
    \label{Figure: all-interactions-9panel-pi-20}
\end{figure}

\begin{figure}
    \centering
    \includegraphics[scale=1.0]{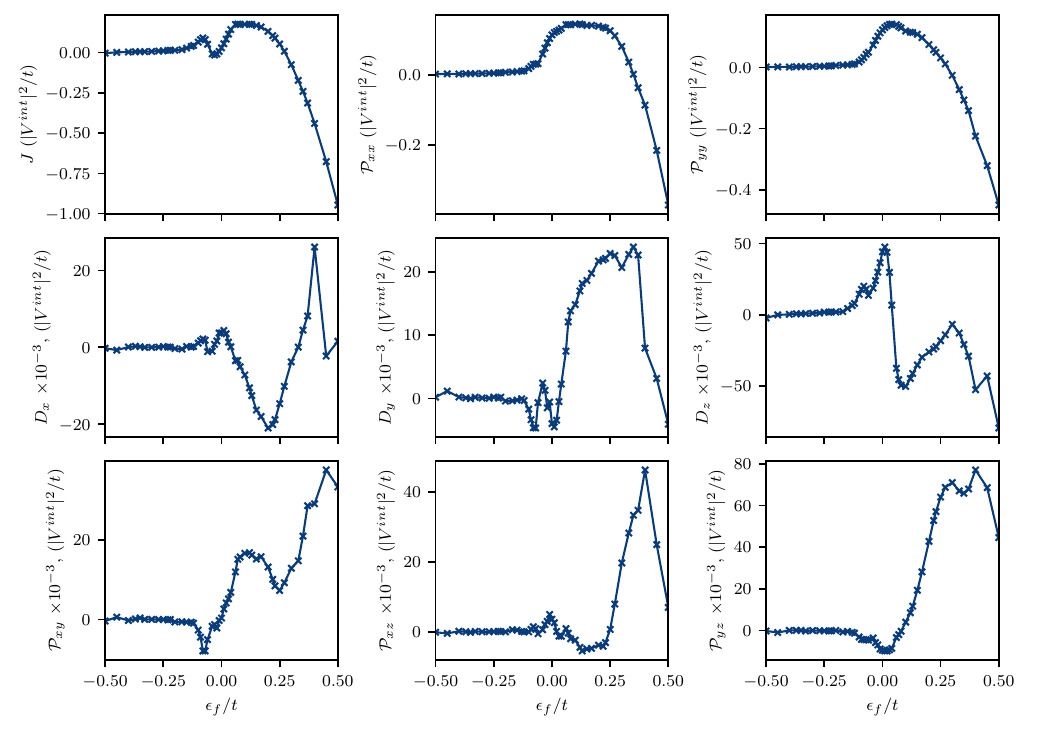}
    \caption{Plots of the interlayer Heisenberg exchange constant $J$, the  DMI vector $\mathbf{D}$, and the five independent components of the traceless, symmetric anistropy matrix $\mathcal{P}$ as functions of the Fermi energy $\epsilon_f$ for the twist angle $\theta = \pi/30$ ($6^{\circ}$). The interlayer hopping is of the Slater-Koster type, $t^{\mathrm{int}}_{\sigma \sigma'}(\mathbf{r}) = V^{\mathrm{int}}  \exp(- r/r_0) \delta_{\sigma \sigma'}$.  The following parameters were used: t = 1 (arb. units), $t_{\uparrow \downarrow} = 0.1t$, $a=\Delta_z = r_0 = 1$ (arb. units).}
    \label{Figure: all-interactions-9panel-pi-30}
\end{figure}

In the final part of this section, we present plots of  $J$, $\mathbf{D}$ and $\mathcal{P}$ for a variety of twist angles $\theta$. Most of the prevalent features in these plots were already discussed in the main body of the paper. Here, we consider only the behavior of the DMI vector $\mathbf{D}$ as the twist angle $\theta$ and the chemical potential $\epsilon_f$ are varied. From Fig. \ref{Figure: all-interactions-9panel-pi-10} to \ref{Figure: all-interactions-9panel-pi-30}, we observe that $\mathbf{D}$, although tending to point primarily in the $z$-direction, can significantly change its orientation by suitable changes in both $\theta$ and $\epsilon_f$. It is thus seen that twist-angle engineering and electrostatic gating/doping not only influence the magnitude of magnetic interaction parameters, but also their direction.

Finally, these plots illustrate the wide tunability of the interlayer magnetic interactions through variations in both twist angle and chemical potential. Depending on the specific combination of these parameters, the relative strength of the Heisenberg exchange interaction, DMI, and anisotropic exchange can be significantly enhanced or suppressed. This highlights the potential of twist-angle engineering and electrostatic control through gating or doping as powerful tools for designing and tuning magnetic properties in van der Waals heterostructures.

\bibliography{biblio}